\documentclass[conference]{IEEEtran}

\usepackage[letterpaper, left=0.74in, right=0.74in, bottom=1.02in, top=0.85in]{geometry}

\usepackage{xcolor}

\makeatletter
\newcommand{\ssymbol}[1]{^{\@fnsymbol{#1}}}
\makeatother

\usepackage[caption=false]{subfig}
\ifCLASSINFOpdf
  \usepackage[pdftex]{graphicx}
\else
\fi
\usepackage{amsmath,amssymb,amsfonts}
\usepackage{amsmath}
\usepackage{textcomp}
\usepackage{algorithm}
\usepackage[noend]{algpseudocode}

\makeatletter
\def\BState{\State\hskip-\ALG@thistlm}
\makeatother

\algnewcommand\algorithmicswitch{\textbf{switch}}
\algnewcommand\algorithmiccase{\textbf{case}}
\algnewcommand\algorithmicassert{\texttt{assert}}
\algnewcommand\Assert[1]{\State \algorithmicassert(#1)}%
\algdef{SE}[SWITCH]{Switch}{EndSwitch}[1]{\algorithmicswitch\ #1\ \algorithmicdo}{\algorithmicend\ \algorithmicswitch}%
\algdef{SE}[CASE]{Case}{EndCase}[1]{\algorithmiccase\ #1}{\algorithmicend\ \algorithmiccase}%
\algtext*{EndSwitch}%
\algtext*{EndCase}%

\usepackage{enumitem}
\usepackage{amsmath}
\usepackage{comment}

\usepackage{float}
\begin{document}
%
\title{Density-Aware~Reinforcement~Learning~to~Optimise Energy Efficiency in UAV-Assisted Networks}


\author{\IEEEauthorblockN{Babatunji Omoniwa\IEEEauthorrefmark{1}, Boris Galkin\IEEEauthorrefmark{2}, Ivana Dusparic\IEEEauthorrefmark{1}}
\IEEEauthorblockA{CONNECT Centre for Future Networks and Communications,\\
\IEEEauthorrefmark{1}Trinity College Dublin, Dublin, Ireland.\\
\IEEEauthorrefmark{2}Tyndall National Institute, Cork, Ireland.\\
Emails: omoniwab@tcd.ie, boris.galkin@tyndall.ie, ivana.dusparic@scss.tcd.ie}
}


%


\maketitle

\begin{abstract}
Unmanned aerial vehicles (UAVs) serving as aerial base stations can be deployed to provide wireless connectivity to mobile users, such as vehicles. However, the density of vehicles on roads often varies spatially and temporally primarily due to mobility and traffic situations in a geographical area, making it difficult to provide ubiquitous service. Moreover, as energy-constrained UAVs hover in the sky while serving mobile users, they may be faced with interference from nearby UAV cells or other access points sharing the same frequency band, thereby impacting the system's energy efficiency (EE). Recent multi-agent reinforcement learning (MARL) approaches applied to optimise the users' coverage worked well in reasonably even densities but might not perform as well in uneven users' distribution, i.e., in urban road networks with uneven concentration of vehicles. In this work, we propose a density-aware communication-enabled multi-agent decentralised double deep Q-network (DACEMAD--DDQN) approach that maximises the total system's EE by jointly optimising the trajectory of each UAV, the number of connected users, and the UAVs' energy consumption while keeping track of dense and uneven users' distribution. Our result outperforms state-of-the-art MARL approaches in terms of EE by as much as 65\% -- 85\%.


\end{abstract}


%

\IEEEpeerreviewmaketitle

\section{Introduction}
Unmanned aerial vehicles (UAVs) have numerous real-world applications, ranging from assisted communication in disaster-affected areas to surveillance, search and rescue operations. In particular, UAVs can be flexibly deployed as base stations in out-of-coverage areas, complementing and lowering the cost of deploying terrestrial infrastructures~\cite{Omoniwa_DQLCI_2021}.
Furthermore, UAVs may be deployed in situations of a sudden increase in mobile users' demand, i.e., network load, or service outage due to disasters~\cite{Omoniwa_DQLCI_2021}--\cite{Mozaffari2017UAV}. With growing interest in the market for connected and autonomous vehicles~\cite{Gueriau2020_dublin} and their requirements for ultra-reliable network connectivity, the reliance on UAVs to provide ubiquitous coverage is expected to sky-rocket.
However, it is challenging to provide coverage in dynamic network environments characterised by the changing density of road vehicles caused by the spatial and temporal variations due to the mobility and traffic situation in a geographical area~\cite{Marini_UAV_V2X_2022}.

To fully benefit from the deployment of UAVs serving mobile users, some major challenges need to be addressed, they include, flight trajectory optimisation~\cite{Omoniwa_DQLCI_2021, omoniwaLetters2022}, energy efficiency (EE) optimisation~\cite{Liu2020UAVdistributed, Galkin2022_ICC_fixed_wings} and coverage optimisation~\cite{Omoniwa_DQLCI_2021}. Specifically, UAVs have limited onboard battery capacity and deplete energy while hovering in the sky and providing coverage for extended periods of time. In addition, multiple UAVs sharing the same frequency spectrum and deployed to provide wireless connectivity to vehicles in a given area may experience a decrease in the EE due to interference from neighbouring UAV cells~\cite{Galkin2022_ICC_fixed_wings}. 

Optimising the EE of UAVs providing coverage to vehicles depends on several factors such as the density of vehicles in the area, the UAVs' energy capacity, the bandwidth requirements, and the communication capability of the UAVs in an interference-limited environment. Several research efforts have been made towards optimising the systems EE while UAVs serve ground users. The work in \cite{Marini_UAV_V2X_2022} presented a meta-reinforcement learning approach to optimise the trajectory of a single UAV while maximising the coverage of vehicles in an urban environment. In this work, we consider the deployment of multiple UAVs providing coverage to vehicles in an urban environment. An iterative approach was proposed in \cite{Mozaffari2017UAV} to optimise the flight trajectory of each UAV such that the total energy used by the UAVs is minimised. However this work only considered static ground devices and relied on a central controller (CC) located at a central cloud server for decision making. A multi-UAV placement problem was presented in \cite{Islam2022_UAV_Vanet} to optimise the coverage of vehicles in an urban area. However, the work relying on a CC may be impractical in a disaster where a possible failure in the CC may lead to a service outage. Moreover, it may be challenging to track users' location in such emergencies. On this note, there has been a shift towards the decentralised control of UAVs, with recent research adopting disruptive machine learning (ML) techniques to solve complex optimisation problems in UAV-assisted networks~\cite{Omoniwa_DQLCI_2021, Liu2020UAVdistributed, omoniwaLetters2022, Galkin2022_ICC_fixed_wings}.

Specifically, reinforcement learning (RL) has been shown to improve the EE of UAVs deployed to serve ground users in dynamic environments~\cite{Liu2020UAVdistributed, Omoniwa2019, Samir2021_UAV_vanet}. A centrally-controlled actor-critic algorithm was proposed in~\cite{Samir2021_UAV_vanet} optimise the trajectories of UAVs while maximising the coverage of vehicles in an interference-free environment. However, as the number of UAVs in the network increases, it becomes impractical for effective decision-making and control in disaster scenarios since a potential loss of control packets to the UAVs may impact the service delivery. The decentralised Multi-Agent Deep Deterministic Policy Gradient (MADDPG) approach proposed in~\cite{Liu2020UAVdistributed} was an improvement to the centralised learning approach in \cite{Liu2018UAV}, where all agents are controlled by a single actor-critic network. Although both work~\cite{Liu2020UAVdistributed} and \cite{Liu2018UAV} focused on optimising the systems' EE while serving static pedestrian users, they did not account for the interference from neighbouring UAV cells. In~\cite{Galkin2022_ICC_fixed_wings}, we proposed a decentralised Multi-Agent Reinforcement Learning (MARL) approach, where each UAV is equipped with a Dueling Deep
Q-Network (DDQN) agent which can adjust the UAV flight
trajectory to optimise the systems' EE. However, the work focused on fixed-winged UAVs providing coverage to static users in rural areas. Our previous work~\cite{omoniwaLetters2022} presented a Multi-Agent Decentralised Double Deep Q-Network (MAD--DDQN) approach to maximise the systems' EE while jointly optimising the trajectory of each UAV, the outage of mobile pedestrians and the energy consumption. However, the absence of direct collaboration among neighbouring UAVs impacted the overall systems' EE.

In our recent work~\cite{omoniwa_cmad_paper}, we proposed a Communication-enabled Multi-Agent Decentralised Double Deep Q-Network (CMAD--DDQN) approach to overcome the shortcomings of the MAD-DDQN approach by supporting direct collaboration among UAVs via a 3GPP-defined communication protocol~\cite{3GPPstandard_neighborRelations} to maximise the systems’ EE. Although the CMAD--DDQN outperforms the MAD-DDQN as the number of UAVs increases, both approaches only worked well in reasonably even densities of geographically-confined users but might not perform as well with an uneven distribution where some areas are denser than others, i.e., in an event scenario with the concentration of users, or mostly in vehicular scenarios where users are congregated in the road space with service fluctuations or outages, in particular congested road space. Based on the identified gaps, this work is motivated towards deploying UAVs to provide wireless connectivity to densely and uneven users in an energy-efficient manner. Hence, we outline our contributions as follows: 
\begin{itemize}[leftmargin=*]
  \item We propose a novel Density-Aware Communication-Enabled Multi-Agent Decentralised Double Deep Q-Network (DACEMAD--DDQN) approach to maximise the systems' EE by jointly optimising the UAVs flight trajectory, the number of connected ground users and the total energy consumed by UAVs in a shared, dynamic and interference-limited environment. Our approach allows for direct collaboration among agent-controlled UAVs to learn policies that maximise the systems' EE while providing coverage to highly mobile and densely uneven users' distribution in real time.
  \item We investigate the effectiveness of our DACEMAD--DDQN approach in intelligently tracking the user density while providing coverage by first testing the algorithm under different ground users configurations. We then consider real traffic data of the Dublin City Centre generated via SUMO~\cite{Gueriau2020_dublin}, where there is a flow of traffic, i.e., the vehicles may enter or leave the coverage region. Furthermore, we assume that the agent-controlled UAVs have no prior knowledge of the locations of vehicles via a CC. The proposed approach outperforms state-of-the-art MARL approaches in maximising the total systems' EE without degrading the coverage performance in the network.
\end{itemize}

\section{System model}
\begin{figure}[t]
\centering
\includegraphics[width=2.5in]{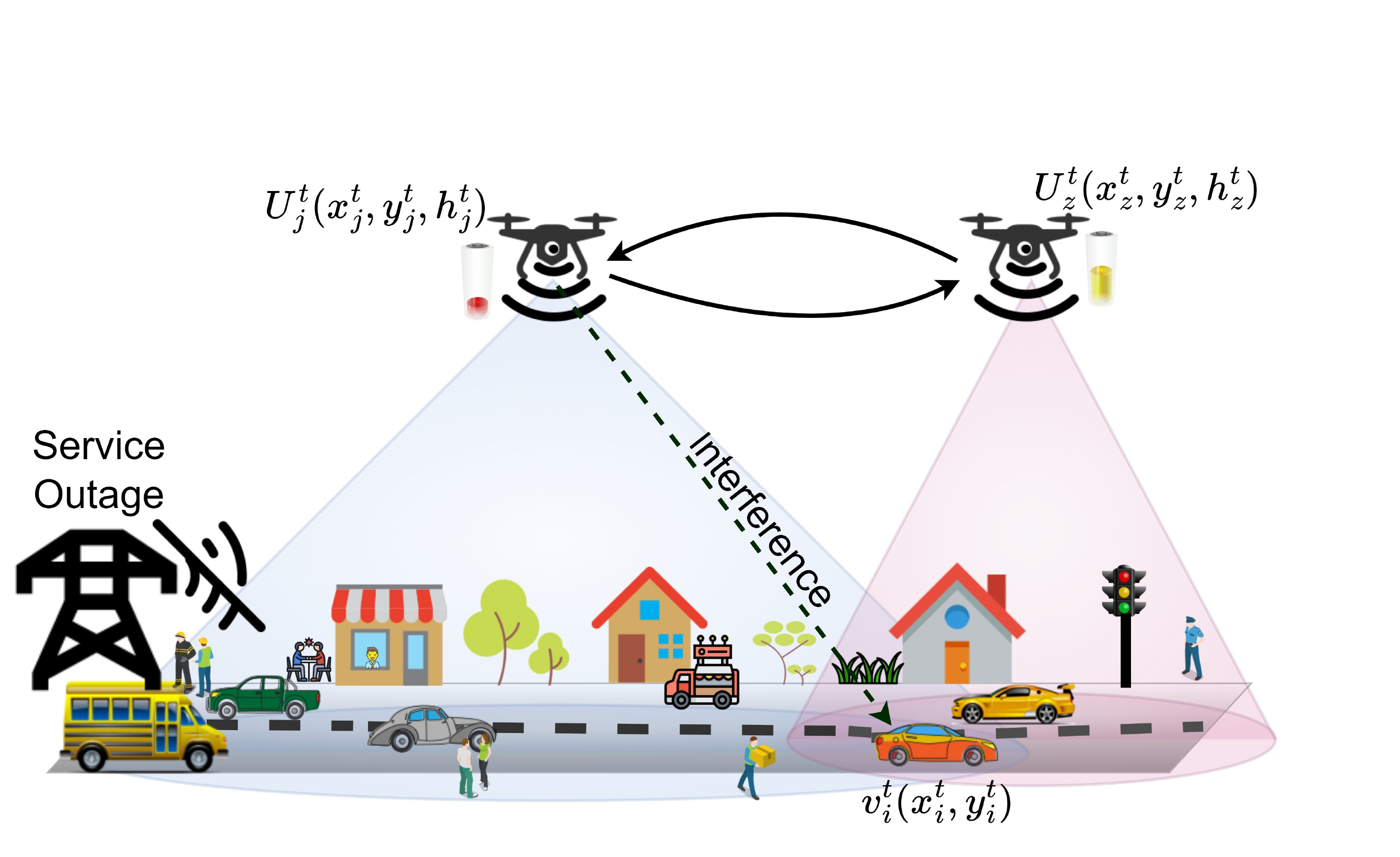}
 \caption{System model for UAVs providing coverage to vehicles.}
 \label{fig:systemmodel}
\end{figure}

\vspace{-0.5mm}
We consider a UAV-assisted vehicular network with a set $U$ of quadrotor UAVs deployed to serve vehicles in an urban setting as shown in Figure \ref{fig:systemmodel}. We assume that each vehicle $i \in v$ is equipped with a transceiver that allows for the transmission and reception of wireless signals. As in \cite{omoniwaLetters2022}, we assume service unavailability in existing terrestrial infrastructure due to disaster, unforeseen load or failure in parts of the network.

\vspace{-1mm}
We assume guaranteed Line-of-Sight conditions between $U_j^t$ located at $(x_j^t, y_j^t, h_j^t)$ and $v_i^t$ at $(x_i^t, y_i^t)$ due to the aerial positions of the UAV. However, the wireless channel is assumed to be impaired by interference from nearby UAV cells or other access points sharing the same frequency spectrum. In time-step $t$, each vehicle $i \in v$ can be served by a single UAV $j \in U$ which provides the strongest downlink signal-to-interference-plus-noise-ratio (SINR). Hence, the SINR at time $t$ is expressed as~\cite{Omoniwa_DQLCI_2021, omoniwaLetters2022},
\begin{equation}\label{eq:sinr}  
    \gamma_{i,j}^t = \beta  P (d_{i,j}^t)^{-\alpha} \Big/\big( \Sigma_{z \in \chi_{int}} \beta P (d_{i,z}^t)^{-\alpha} + \sigma^2\big),
\end{equation}
where $\beta$ and $\alpha$ are the attenuation factor and path loss exponent that characterises the wireless channel, respectively. $\sigma^2$ is the power of the additive white Gaussian noise at the receiver, $d_{i,j}^t$ is the distance between the vehicle $i$ and UAV $j$ at time $t$.~$\chi_{int} \in U$ is the set of interfering UAVs. $z$ is the index of an interfering UAV in the set~$\chi_{int}$. The transmit power of the UAV is denoted as $P$. To provide ubiquitous connectivity to the vehicles, the UAVs must optimise their flight trajectories. Given a channel bandwidth $B_w$, the receiving data rate at the vehicle can be expressed using Shannon's equation~\cite{Galkin2022_ICC_fixed_wings},
\begin{equation}\label{eq:shannon}
  \mathbb{R}_{i,j}^t = B_w\log_2(1 + \gamma_{i,j}^t).  
\end{equation}

In our interference-limited system, coverage is affected by the SINR. Hence, we compute the connectivity score of a UAV $j \in N$ at time $t$ as~\cite{Liu2020UAVdistributed},
\begin{equation}\label{eq:avgcoveragescore}
C_j^t = \sum_{\forall i \in v}w_j^t(i),
\end{equation}
where $w_j^t(i) \in [0, 1]$ denotes whether vehicle $i$ is connected to UAV $j$ at time $t$. $w_j^t(i) = 1$ if $\gamma_{i}^t > \gamma_{th}$, otherwise $w_j^t(i) = 0$, where $\gamma_{th}$ is the SINR predefined threshold. Likewise $\mathbb{R}_{i,j}^t = 0$ if vehicle $i$ is not connected to UAV $j$.\\ 
We consider the propulsion power consumption model for a rotary-wing UAV used in \cite{omoniwaLetters2022}. A closed-form analytical propulsion power consumption model for a rotary-wing UAV at time $t$ is given as \cite{Zeng2019UAV},
\vspace{-2mm}
\begin{equation}\label{eq:powertime}
P(t) = \kappa_0 \Big(1 + \frac{3V^2}{U^2_{tip}}\Big)  + \kappa_1 \Big( \sqrt{1 + \frac{V^4}{4v_0^4}} + \frac{V^2}{2v_0^2}\Big)^{\frac{1}{2}} + \frac{\kappa_2}{2} V^3,
\end{equation}
where $\kappa_0$, $\kappa_1$ and $\kappa_2$ are the UAVs' flight constants (e.g., rotor radius, disk area, drag ratio, air density, weight), $U_{tip}$ is the rotor blade's tip speed, $v_0$ is the mean hovering velocity, and $V$ is the UAVs' speed at time $t$. In particular, we take into account the basic operations of the UAV, such as, hovering and acceleration. In particular, we take into account the basic operations of the UAV, such as, hovering and acceleration. During flight operations, the total energy consumed by UAV $j$ at time $t$ is given as~\cite{Omoniwa_DQLCI_2021},
\vspace{-1mm}
\begin{equation}\label{eq:propenergy}
e_{j}^t = \delta_t \cdot P(t),
\vspace{-1mm}
\end{equation}
where $\delta_t$ is the duration of each time-step. The EE of UAV $j$ can be expressed as the ratio of the data throughput and the energy consumed in time-step $t$. Therefore, the total systems' EE over all time-step is given as,
\begin{equation}\label{eq:tot_energyEfficiency}
    \eta_{tot} =~ \sum\limits_{t=1}^T \sum\limits_{j \in U} \sum\limits_{i \in v} \mathbb{R}_{i,j}^t \Bigg/ \Bigg(\sum\limits_{t=1}^T \sum\limits_{j \in U} e_j^t \Bigg).   
\end{equation}

\section{Multi-Agent Reinforcement Learning Approach for Energy Efficiency Optimisation}
\label{section:DACE}
In this section, we formulate the problem and propose a DACEMAD--DDQN algorithm to improve the trajectory of each UAV in a manner that maximises the total system's EE.

\subsection{Problem Formulation}
Our objective is to maximise the total system's EE by jointly optimising each UAV's trajectory, number of connected vehicles, and the energy consumed by the UAVs under a strict energy budget. Therefore, the problem is formulated as, 
\begin{subequations}
\begin{align} \label{eq:problem_statement}
&\underset{\forall j \in U:~\mathbf{x_j^t,~y_j^t,~e_j^t,~C_j^t}}{\max}~\eta_{tot}\\
\label{eq:problem_statement_b}
\text{s.t.} \quad &  \gamma_{i,j}^t \geq \gamma_{th},\quad \forall w_j^t(i) \in [0, 1],~i,~j,~t, \\ \label{eq:problem_statement_c}
&e_j^t \leq e_{\max},\quad \quad \quad \quad \quad \quad \quad \forall j,~t,\\ \label{eq:problem_statement_d}
&x_{\min}~\leq~x_j^t~\leq x_{\max},\quad \quad ~~\forall j,~t,\\ \label{eq:problem_statement_e}
&y_{\min}~\leq~y_j^t~\leq y_{\max},\quad \quad ~~~\forall j,~t,
\end{align}
\end{subequations}
where $x_{min}$, $y_{min}$ and $x_{max}$, $y_{max}$, are the UAVs' minimum and maximum coordinates of $x$ and $y$, respectively. $e_{\max}$ is the UAV's maximum energy budget. The constraints in (\ref{eq:problem_statement_b})--(\ref{eq:problem_statement_e}) ensure that the UAVs stay within tolerable bounds. As multiple wireless transmitters sharing the same frequency spectrum are deployed in close proximity to each other, it becomes more challenging to manage interference in the network. The problem (\ref{eq:problem_statement}) is non-convex, thus having multiple local optima. In particular, the problem in (\ref{eq:problem_statement}) is known to be NP-hard~\cite{omoniwaLetters2022}.
Hence, it is intractable to solve using conventional optimization approaches~\cite{Mozaffari2017UAV}. Furthermore, the non-stationarity introduced in the environment results in selfish behaviours in UAVs making them seek individual goals rather than collective goals. As such, it becomes imperative to investigate cooperative strategies that will improve the total system's EE while completing the coverage tasks under dynamic settings.

\begin{figure}[t]
\centering
\includegraphics[width=2.8in]{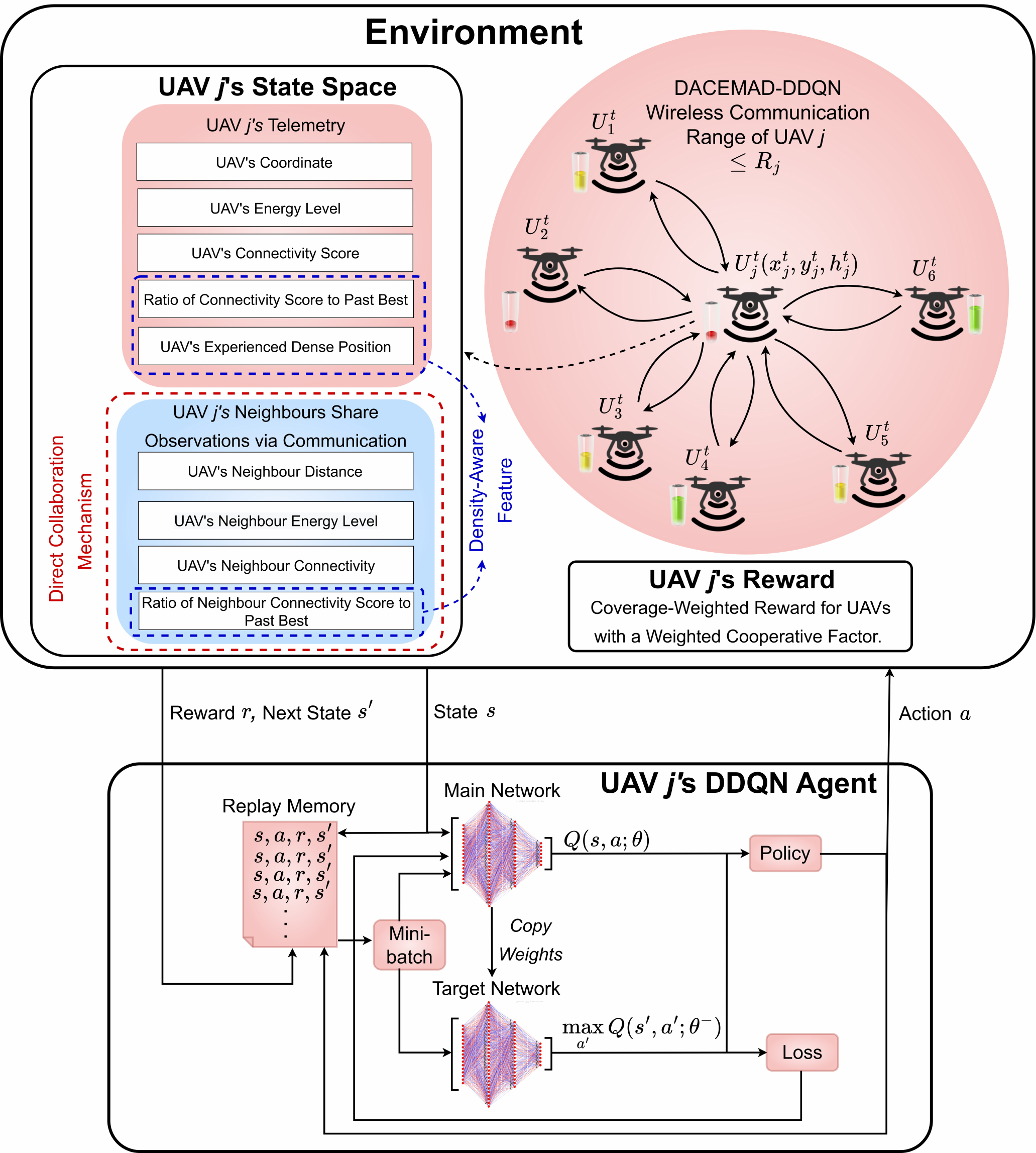}
 \caption{Density-Aware Communication-Enabled Multi-Agent Decentralised Double Deep Q-Network (DACEMAD--DDQN) framework where each UAV $j$ equipped with a DDQN agent interacts with its environment via collaboration and sharing some state information with neighbours within its communication range.} 
 \label{fig: dacemad}
\end{figure}
\begin{algorithm}[ht]
\scriptsize
\caption{Density-Aware Communication-Enabled MAD--DDQN for Agent $j$}\label{Algorithm}
\begin{algorithmic}[1]
\State \textbf{Input:} UAV3Dposition $(x_j^t,~y_j^t,~h_j^t)$, $c^t_j$, InstantaneousEnergyConsumed $e^t_j$, $\frac{c^t_j}{c^*_j}$, ExperiencedDensePosition $(x_j^*,~y_j^*)$, UAVneighbourDistances $N_d^t$, NeighboursConnectionScore $C_z^t$, $\frac{c^t_o}{c^*_o}$, NeighboursInstantaneousEnergyConsumed $e_z^t$ $\in S$ and Output: Q-values corresponding to each possible action~$(+x_s,~0)$, $(-x_s,~0)$, $(0,+y_s)$, $(0,-y_s)$, $(0,~0)$~$\in A_j$. Given the ConnectivityScore $c^t_j$, PastBestConnectivityScore $c^*_j$, NeighbourConnectivityScore $c^t_o$, BestNeighbourConnectivityScore $c^*_o$ .
\ForAll{$a \in A_j$ and~$s \in S$}:
\State \parbox[t]{0.8\linewidth}{$Q_{(1)}(s, a)$, $Q_{(2)}(s, a)$, $\mathcal{D}$ -- empty replay buffer,  $\theta$ -- initial network parameters, $\theta^{-}$ -- copy of $\theta$, $N_r$ -- maximum size of replay buffer, $N_b$ -- batch size, $N^{-}$ -- target replacement frequency.}
\State $s$ $\gets$ initial state
\State 1500 $\gets$ maxStep
\While{\emph{goal} not Reached and Agent \emph{alive} and maxStep not reached} 
\State \emph{s} $\leftarrow$ MapLocalObservationToState(\emph{Env}) 
\State \Comment{\parbox[t]{0.8\linewidth}{\texttt{Execute $\epsilon$-greedy method based on $\pi_j$}}}
\State \emph{a} $\leftarrow$ DeepQnetwork.SelectAction(\emph{s})
\State \Comment{\parbox[t]{0.8\linewidth}{\texttt{Agent executes action in state $s$}}}
\State  \emph{a}.execute(\emph{Env}) \label{Algline:action_execute}
\If { \emph{a}.execute(\emph{Env}) is True}
\State \Comment{\parbox[t]{0.8\linewidth}{\texttt{Map observations to new state $s'$}}}
\State Env.UAVposition  \label{Algline:state_start}
\State Env.ConnectivityScore
\State Env.InstantaneousEnergyConsumed
\State Env.RatioOfConnectivityScore\\
~~~~~~~~~~~~~~~~~~ToPastBestConnectivityScore
\State Env.ExperiencedDensePosition
\State \Comment{\parbox[t]{0.8\linewidth}{\texttt{Map communicated observations from closest neighbours based on an existing ANR mechanism for UAV communication to new state $s'$}}}
\State Env.Neighbour.UAVneighbourDistances
\State Env.Neighbour.ConnectivityScore 
\State Env.Neighbour.RatioOfNeighborhoodConnectivity\\  ~~~~~~~~~~~~~~~~~~ScoreToPastBestNeighborhoodConnectivityScore
\State Env.Neighbour.InstantaneousEnergyConsumed \label{Algline:state_stop}
\EndIf
\State \emph{r} $\leftarrow$ Env.RewardWithCooperativeNeighbourFactor (\ref{eqnreward}) \label{Algline:reward}
\State \textbf{update} $(x_j^*,~y_j^*)$, $c^*_j$, $c^*_o$  $\forall$ $t$
\If {$c^t_j$ $>$ $c^*_j$} \label{Algline:uav_dense_start}
\State $(x_j^*,~y_j^*)$ $\leftarrow$ $(x_j^t,~y_j^t)$
\State $c^*_j$ $\leftarrow$ $c^t_j$  \EndIf \label{Algline:uav_dense_stop}
\If {$c^t_o$ $>$ $c^*_o$} \label{Algline:neighbour_dense_start}
\State $c^*_o$ $\leftarrow$ $c^t_o$ 
\EndIf \label{Algline:neighbour_dense_stop}
\State \Comment{\parbox[t]{0.8\linewidth}{\texttt{Execute UpdateDDQNprocedure()}}}
\State Sample minibatch of $N_b$ tuples $(s, a, r, s') \sim Unif(\mathcal{D})$ \label{Algline:ddqn_start}
\State Construct target values, one for each of the $N_b$ tuples:
\State Define $a^{max} (s'; \theta) =  \arg \max_{a'} Q_{(1)}(s', a'; \theta)$
\If {\emph{$s'$} is Terminal}
\State $y_j =  r$
\Else
\State $y_j =  r + \gamma Q_{(2)}(s', a^{max} ((s'; \theta); \theta^{-})$
\EndIf
\State Apply gradient descent step with loss $\parallel y_j - Q(s, a; \theta)\parallel^2$
\State Replace target parameters $\theta^{-} \gets \theta$ every $N^{-}$ step \label{Algline:ddqn_stop}
\EndWhile
\State \textbf{endwhile}
\EndFor
\end{algorithmic}
\end{algorithm}
\subsection{Density-Aware Communication-Enabled Multi-Agent Decentralised Double Deep Q-Network (DACEMAD--DDQN)}
We assume that each UAV is controlled by a Double Deep Q-Network (DDQN) agent which can learn the density of vehicles in the network, and then adjust its trajectory in such a way that will maximise the total system's EE while jointly optimising the total number of connected vehicles and the energy utilisation of the UAV. Nevertheless, in a typical multi-agent setting, it is often hard to achieve cooperation~\cite{Dafoe_Cooperative_AI} since the interference-limited environment pushes agents to exhibit some selfish behaviors~\cite{omoniwaLetters2022}. Therefore, a robust and adaptive strategy is required to allow agents to collaborate while completing their tasks.

Algorithm \ref{Algorithm} shows the DACEMAD--DDQN for Agent $j$. The DACEMAD--DDQN approach extends the CMAD--DDQN~\cite{omoniwa_cmad_paper} approach, which relies on a communication mechanism based on the existing 3GPP standard~\cite{3GPPstandard_neighborRelations}. However, the DACEMAD--DDQN approach equips each agent with the knowledge of the number of connected vehicles in its neighbourhood and keeps track of its best-experienced coverage during the training phase. From Algorithm~\ref{Algorithm}, Agent $j$ follows an $\epsilon$--greedy policy by executing an action $a$ (line~\ref{Algline:action_execute}), transiting from state $s$ (line \ref{Algline:state_start}--\ref{Algline:state_stop}) to a new state $s'$ and receiving a reward (line \ref{Algline:reward}) given in (\ref{eqnreward}). At each time-step during the training phase, each agent keeps track of its best-experienced connectivity score and also keeps track of that position where it experienced the best number of connected vehicles/users as shown on line~\ref{Algline:uav_dense_start}--\ref{Algline:uav_dense_stop}. Furthermore, each agent keeps track of the best-experienced connectivity score in its neighbourhood as shown on line~\ref{Algline:neighbour_dense_start}--\ref{Algline:neighbour_dense_stop}, which is achieved via communicating with its closest neighbours. The DDQN procedure described on line \ref{Algline:ddqn_start}--\ref{Algline:ddqn_stop} optimises the agent's decisions. To optimise the UAVs' trajectory towards serving densely and uneven users' distribution, we design the state space, action space and reward function as follows:

\begin{figure*}[!htbp]
\begin{minipage}[b]{0.33\linewidth}
\centering
\subfloat[Subfigure 1 list of figures text][Simulation scenario 1 at 10$^{th}$ episode.]{\includegraphics[width=\textwidth]{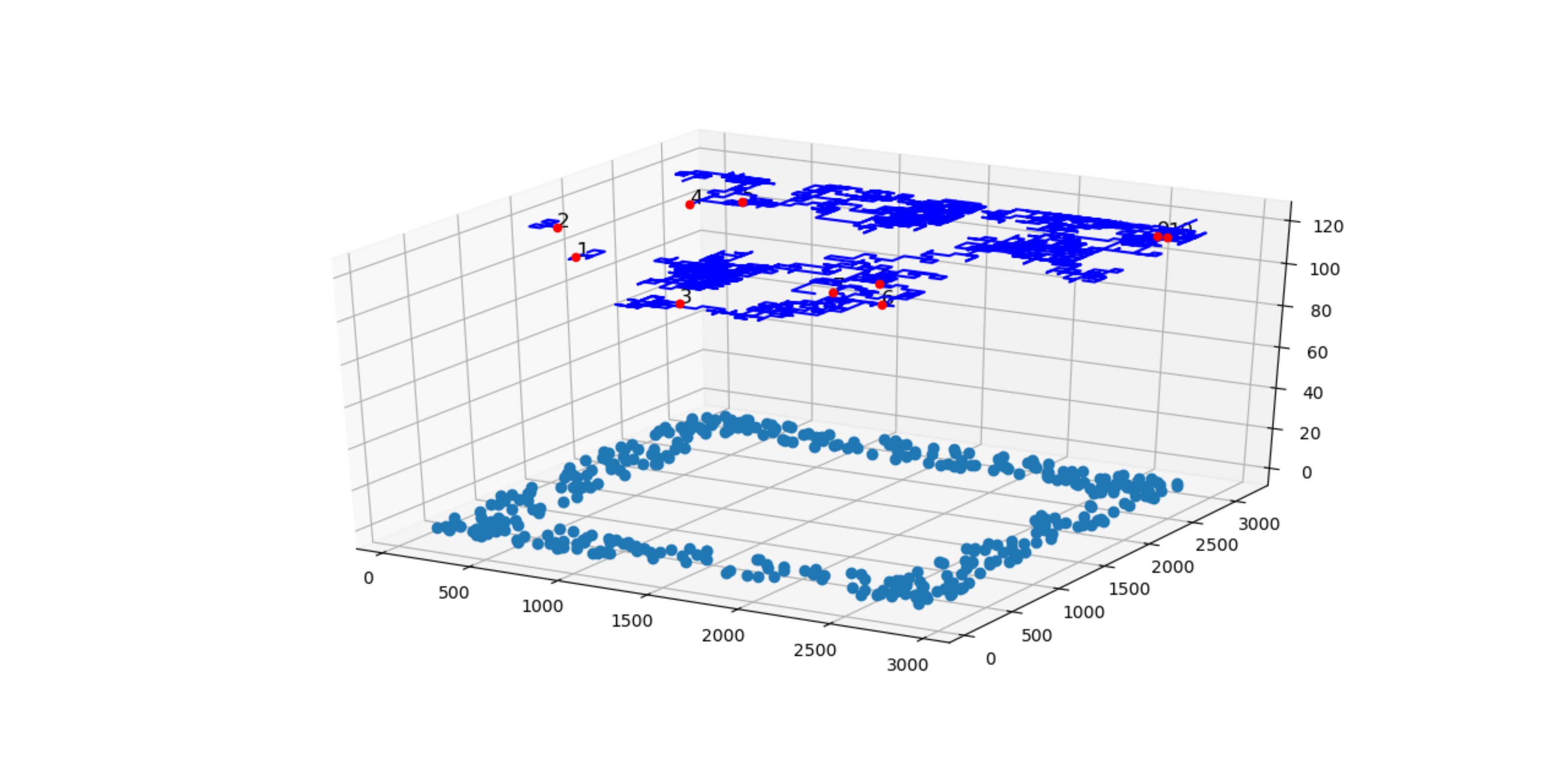}
\label{subfig:circular10}}
\end{minipage}
\begin{minipage}[b]{0.33\linewidth}
\centering
\subfloat[Subfigure 2 list of figures text][Simulation scenario 2 at 10$^{th}$ episode.]{
\includegraphics[width=\textwidth]{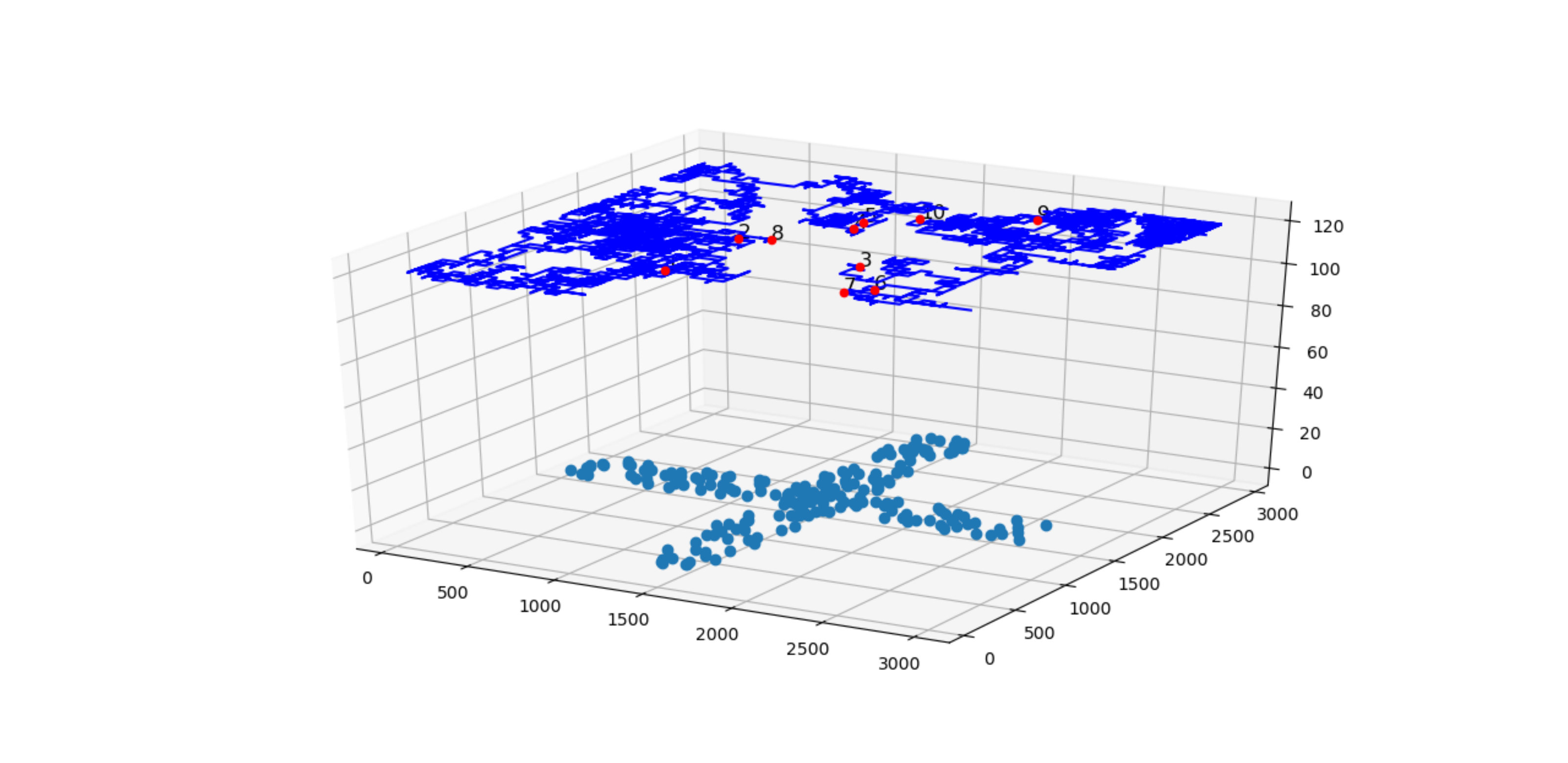}
\label{subfig:cross10}}
\end{minipage}
\begin{minipage}[b]{0.33\linewidth}
\centering
\subfloat[Subfigure 3 list of figures text][Simulation scenario 3 at 10$^{th}$ episode.]{
\includegraphics[width=\textwidth]{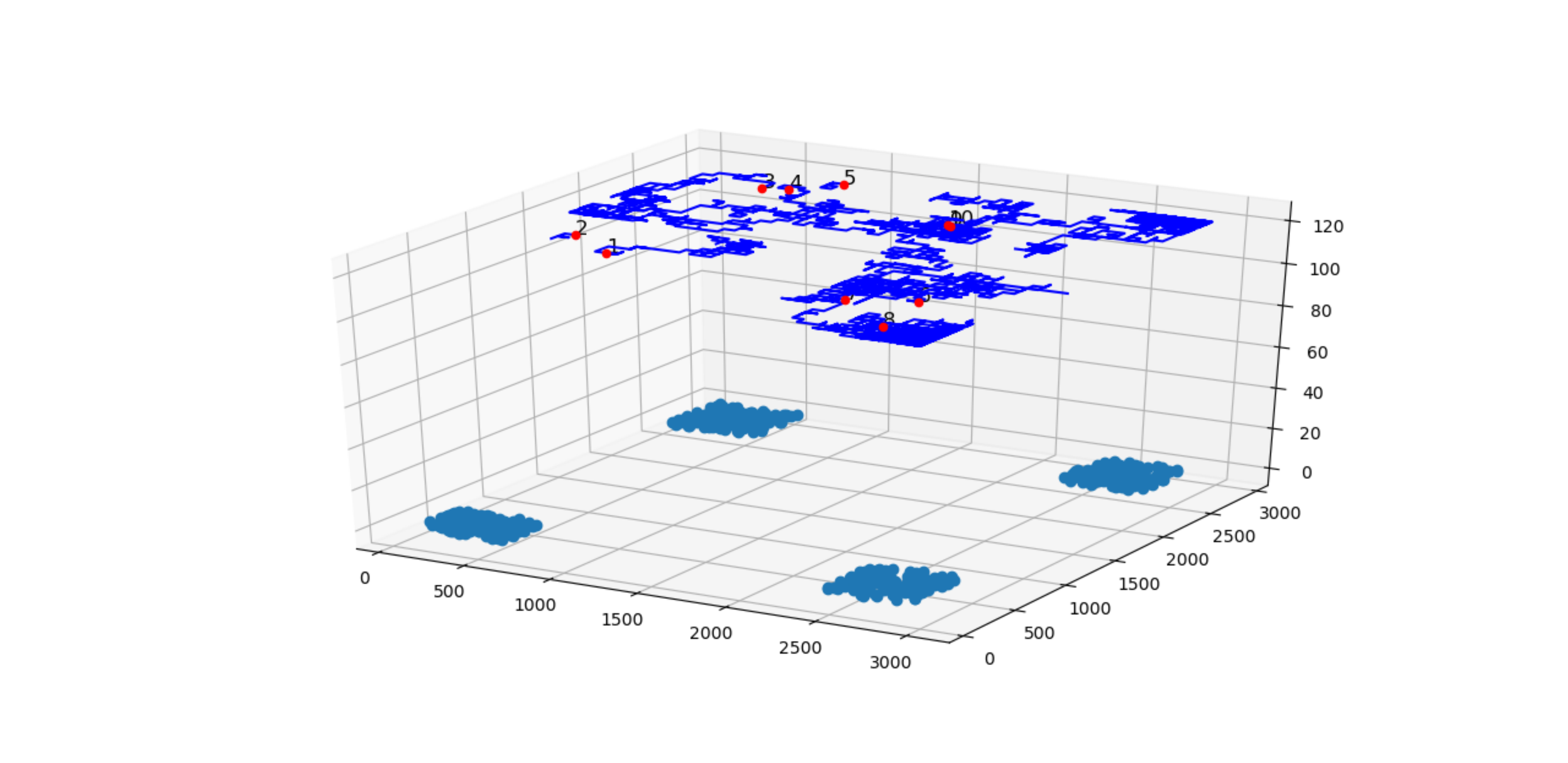}
\label{subfig:edge10}}
\end{minipage}
\begin{minipage}[b]{0.33\linewidth}
\centering
\subfloat[Subfigure 1 list of figures text][Simulation scenario 1 at 250$^{th}$ episode.]{\includegraphics[width=\textwidth]{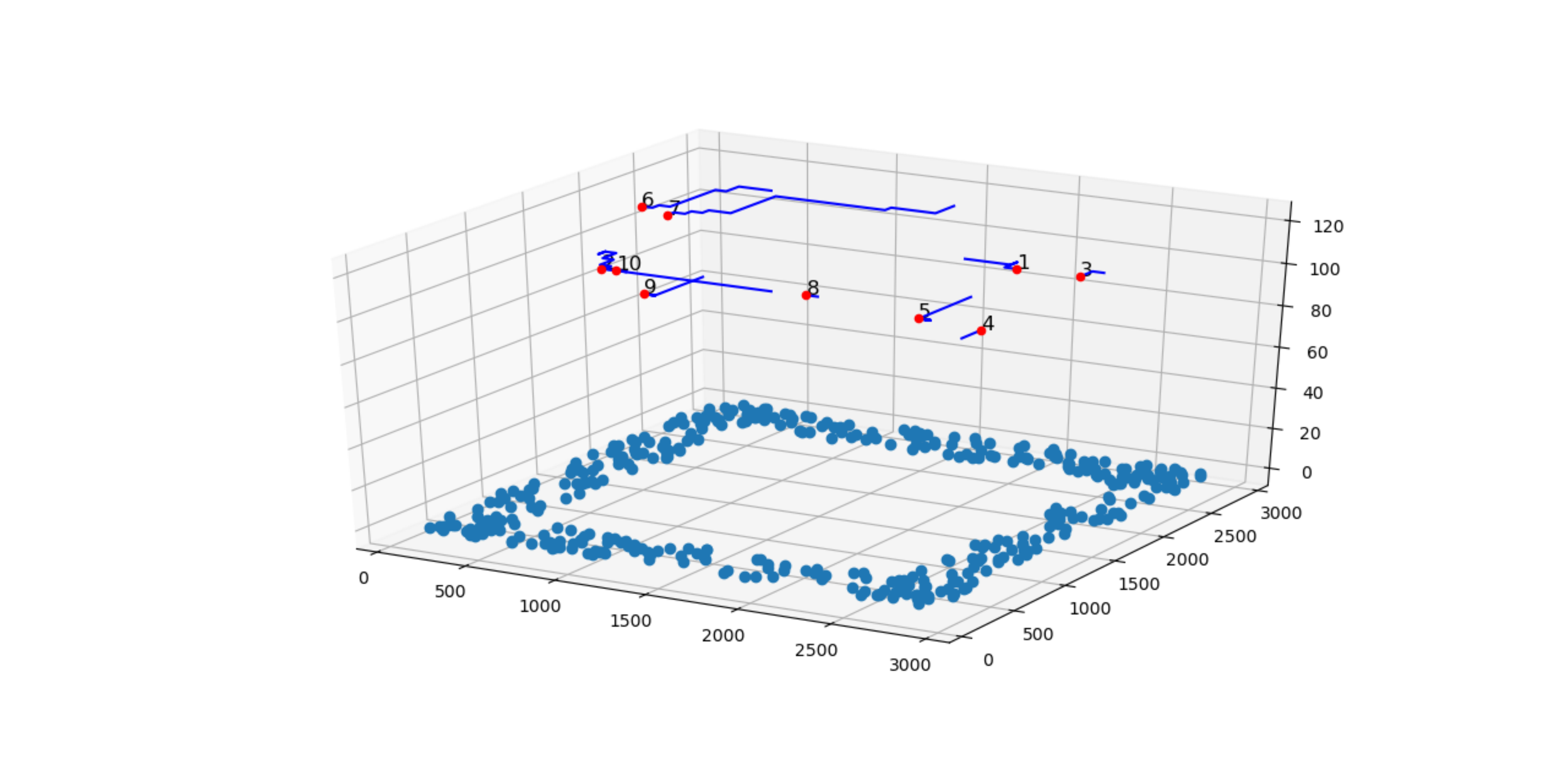}
\label{subfig:circular}}
\end{minipage}
\begin{minipage}[b]{0.33\linewidth}
\centering
\subfloat[Subfigure 2 list of figures text][Simulation scenario 2 at 250$^{th}$ episode.]{
\includegraphics[width=\textwidth]{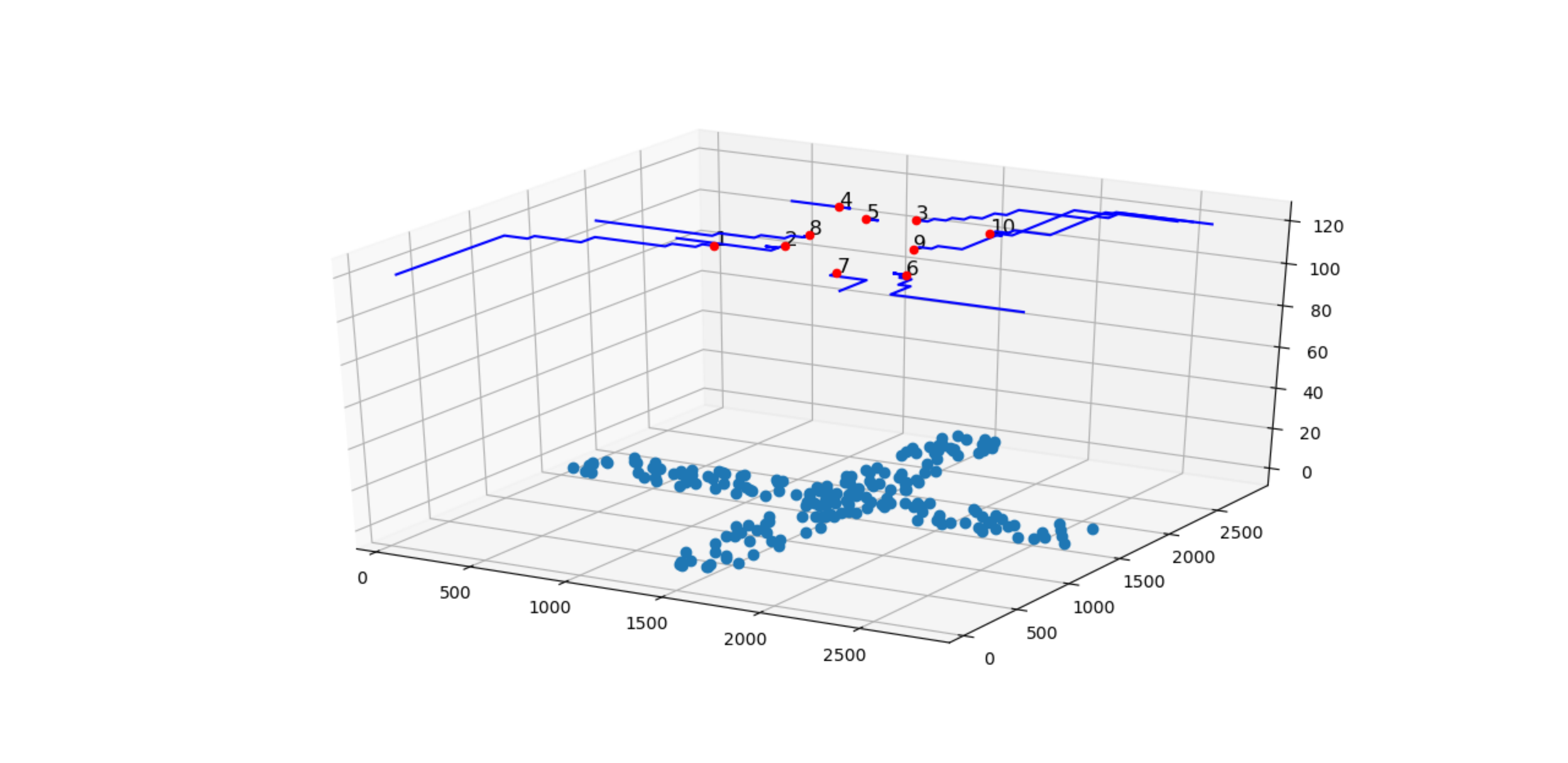}
\label{subfig:cross}}
\end{minipage}
\begin{minipage}[b]{0.33\linewidth}
\centering
\subfloat[Subfigure 3 list of figures text][Simulation scenario 3 at 250$^{th}$ episode.]{
\includegraphics[width=\textwidth]{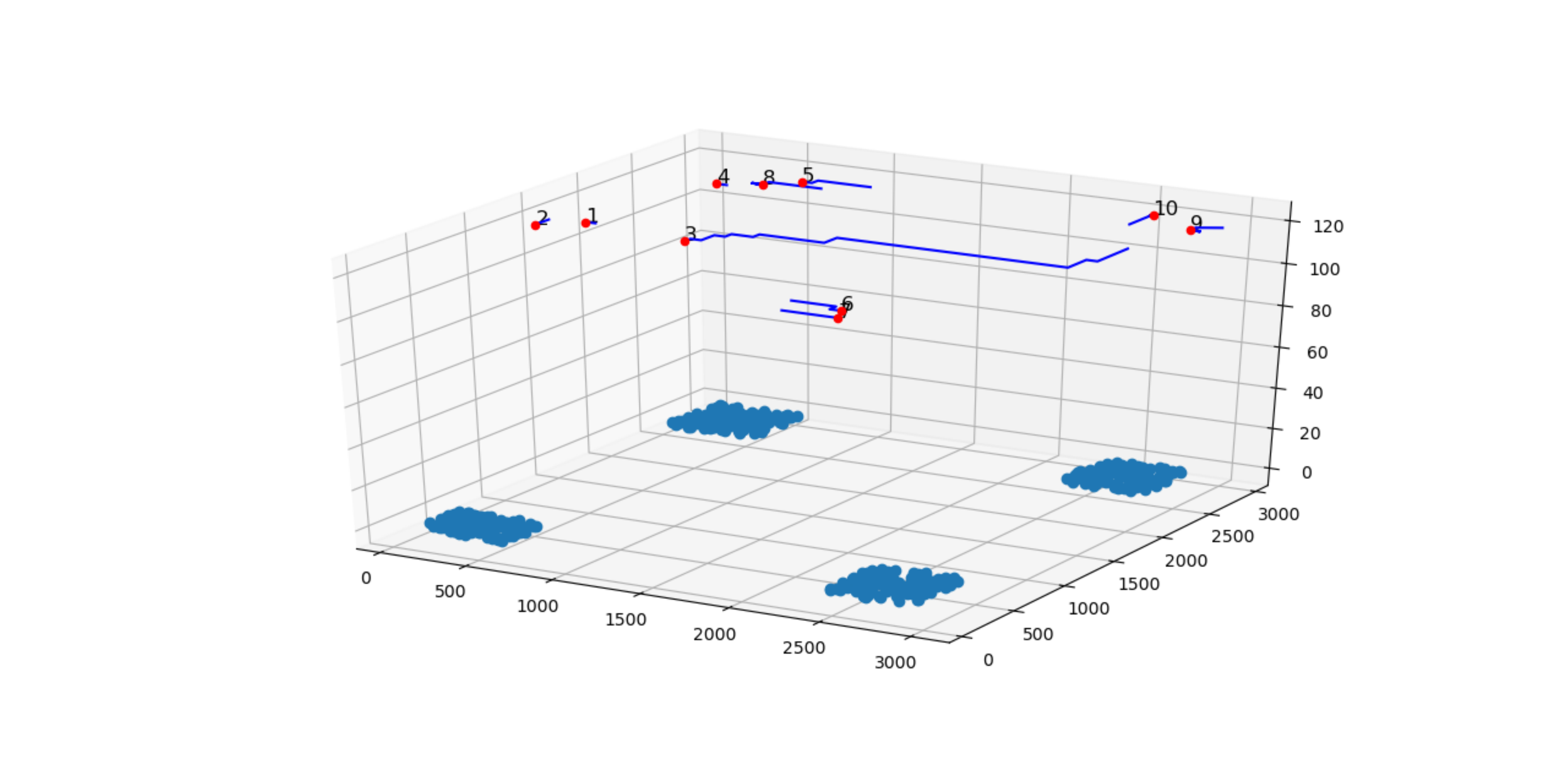}
\label{subfig:edge}}
\end{minipage}
\begin{minipage}[b]{0.33\linewidth}
\centering
\subfloat[Subfigure 1 list of figures text][Top view of scenario 1 at 250$^{th}$ episode.]{\includegraphics[width=\textwidth]{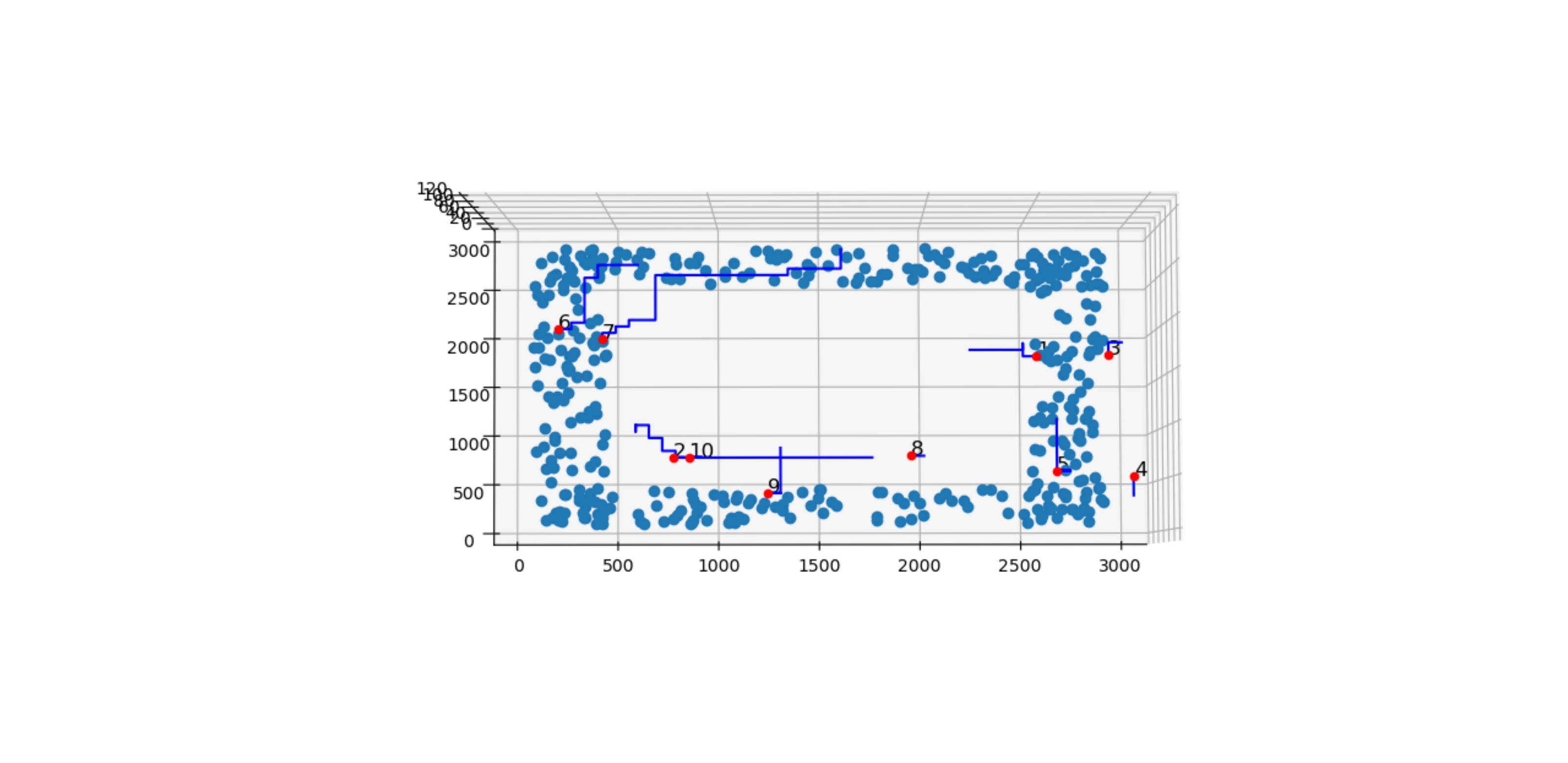}
\label{subfig:circular_top}}
\end{minipage}
\begin{minipage}[b]{0.33\linewidth}
\centering
\subfloat[Subfigure 2 list of figures text][Top view of scenario 2 at 250$^{th}$ episode.]{
\includegraphics[width=\textwidth]{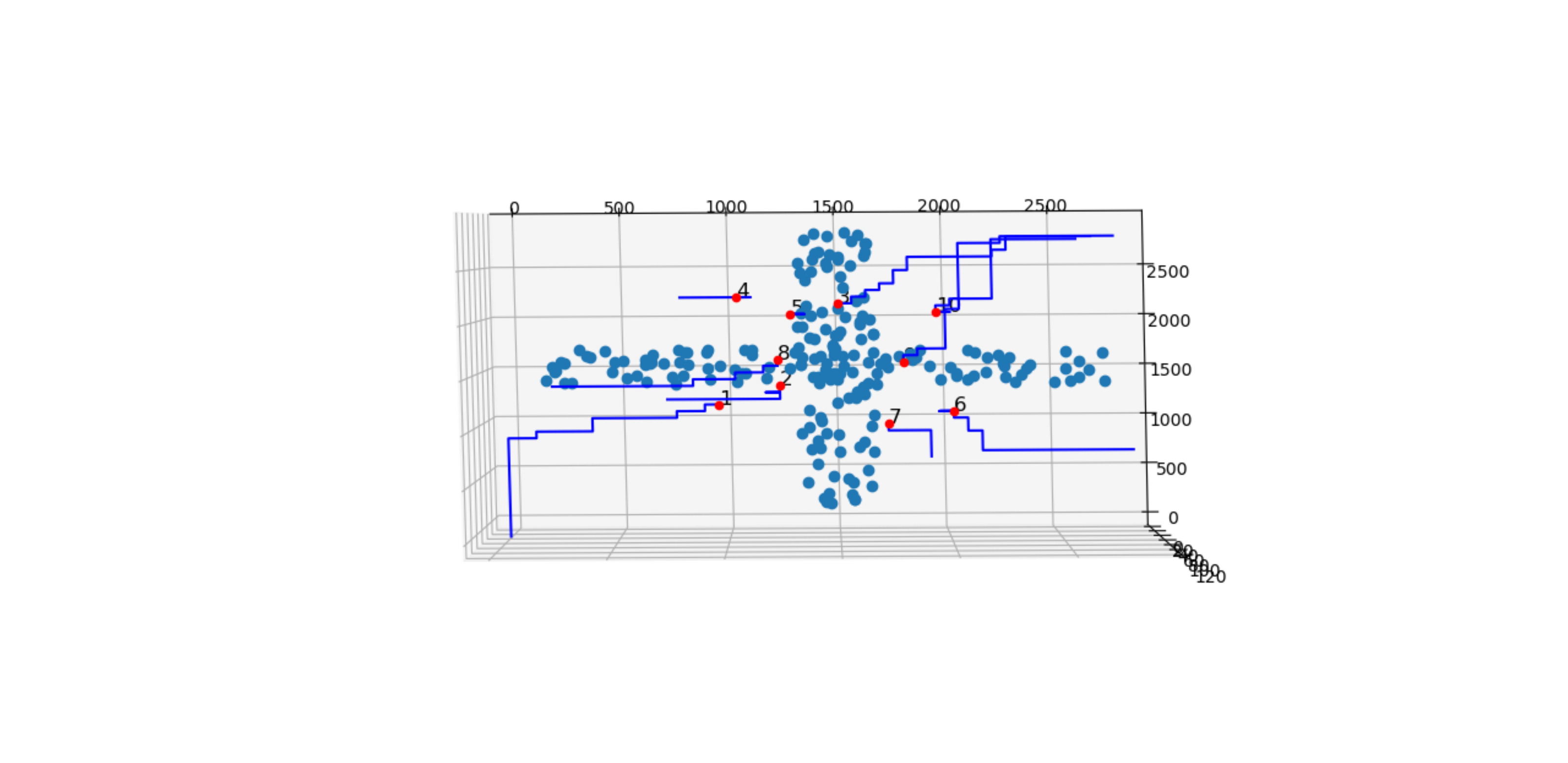}
\label{subfig:cross_top}}
\end{minipage}
\begin{minipage}[b]{0.33\linewidth}
\centering
\subfloat[Subfigure 3 list of figures text][Top view of scenario 3 at 250$^{th}$ episode.]{
\includegraphics[width=\textwidth]{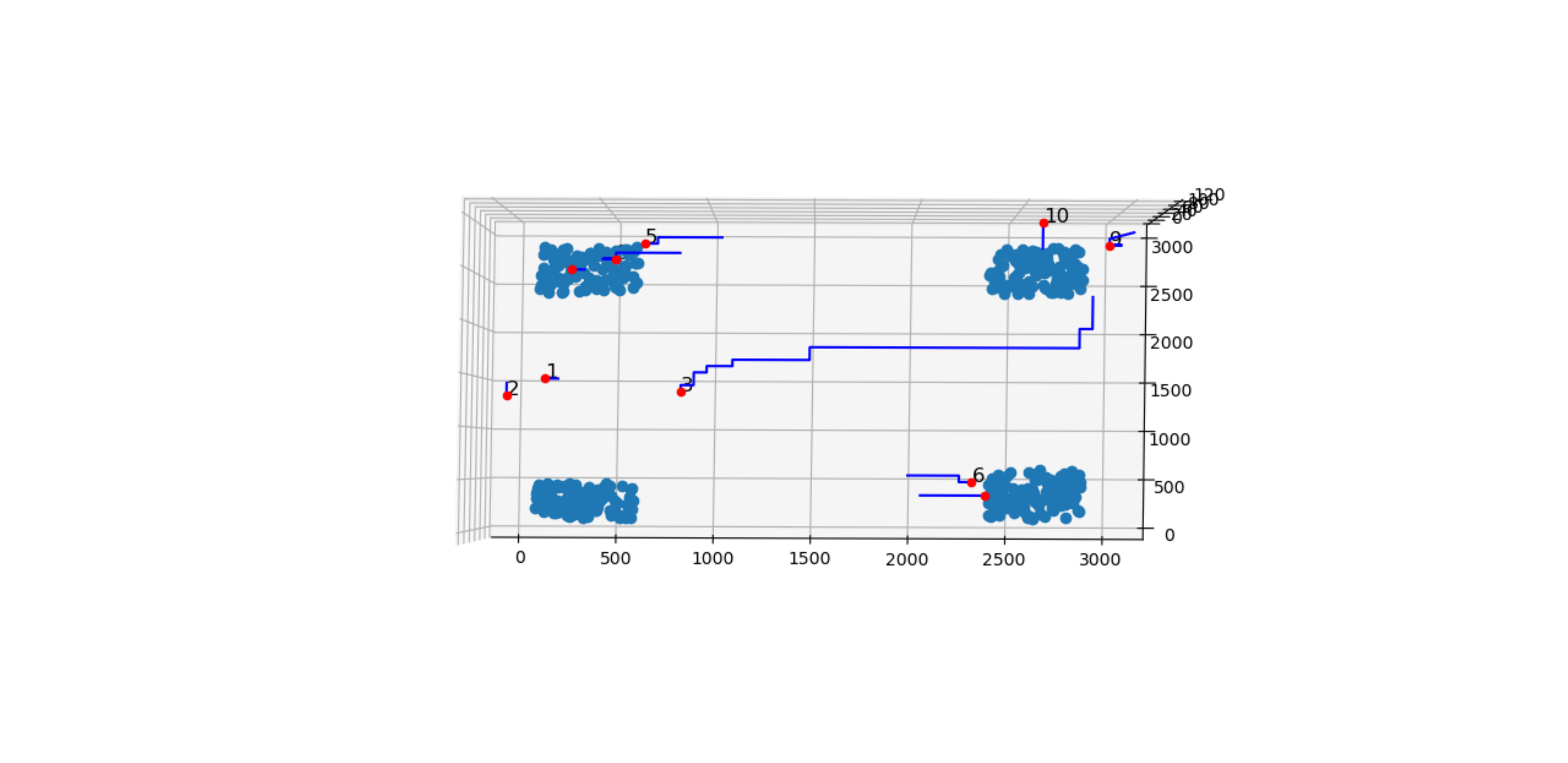}
\label{subfig:edge_top}}
\end{minipage}
\hspace{0.1cm}
\begin{minipage}[b]{0.32\linewidth}
\centering
\subfloat[Subfigure 1 list of figures text][Scenario 1's connected users to deployed users ratio (CDR) vs. episodes.]{\includegraphics[width=0.9\textwidth]{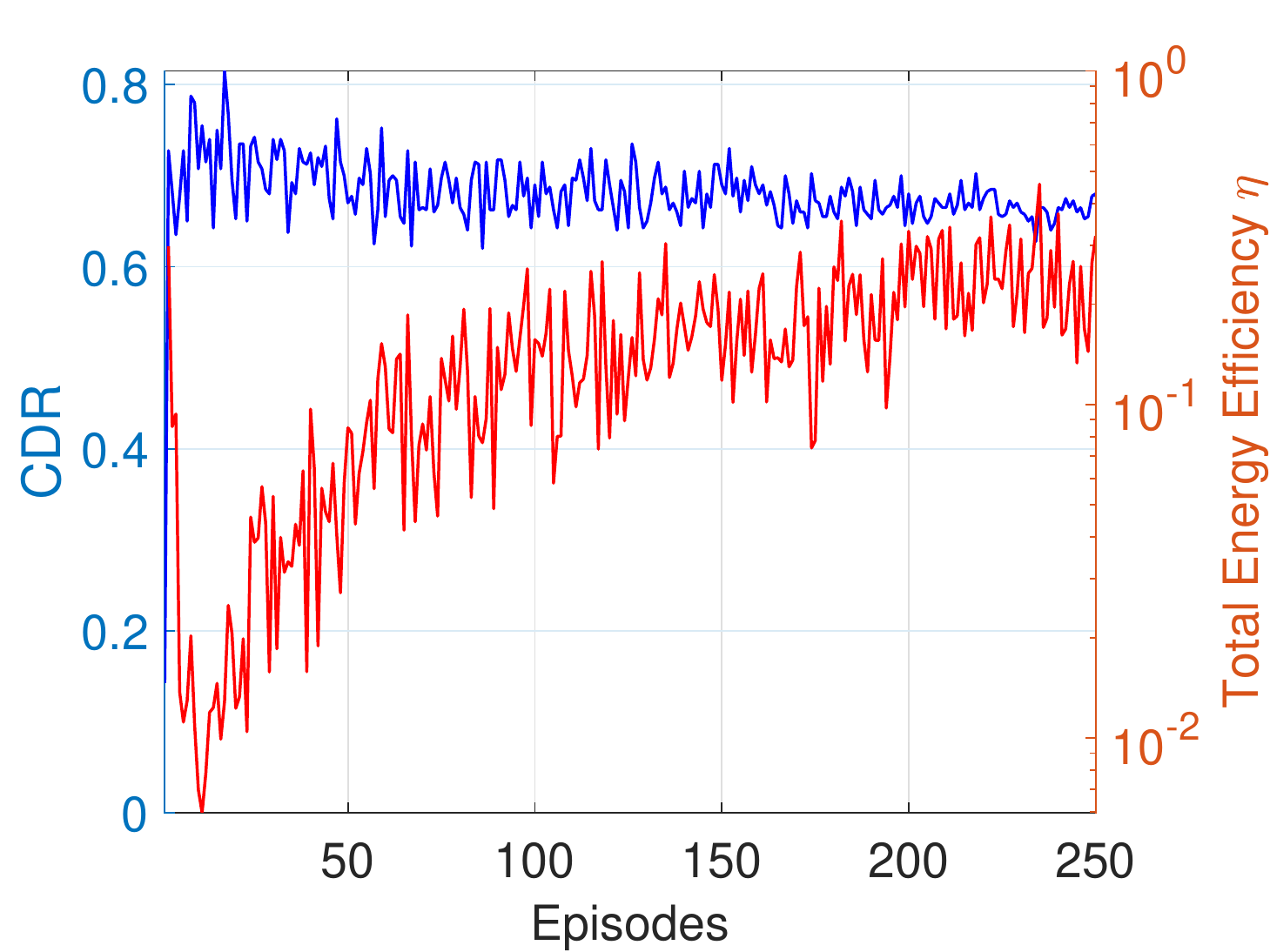}
\label{subfig:circular_cov}}
\end{minipage}
\hspace{0.3cm}
\begin{minipage}[b]{0.32\linewidth}
\centering
\subfloat[Subfigure 2 list of figures text][Scenario 2's connected users to deployed users ratio (CDR) vs. episodes.]{
\includegraphics[width=0.9\textwidth]{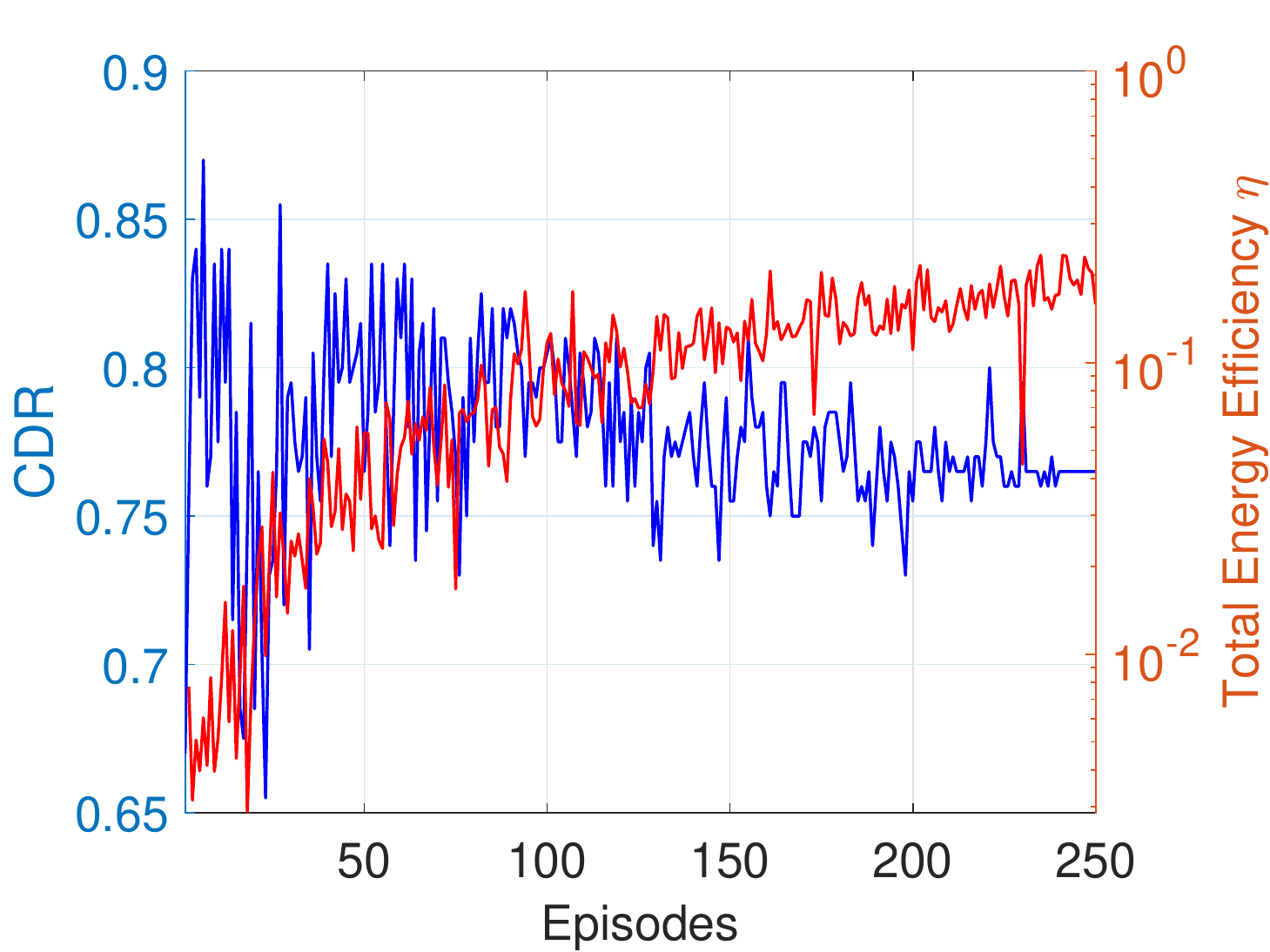}
\label{subfig:cross_cov}}
\end{minipage}
\hspace{0.3cm}
\begin{minipage}[b]{0.32\linewidth}
\centering
\subfloat[Subfigure 2a list of figures text][Scenario 3's connected users to deployed users ratio (CDR) vs. episodes.]{
\includegraphics[width=0.9\textwidth]{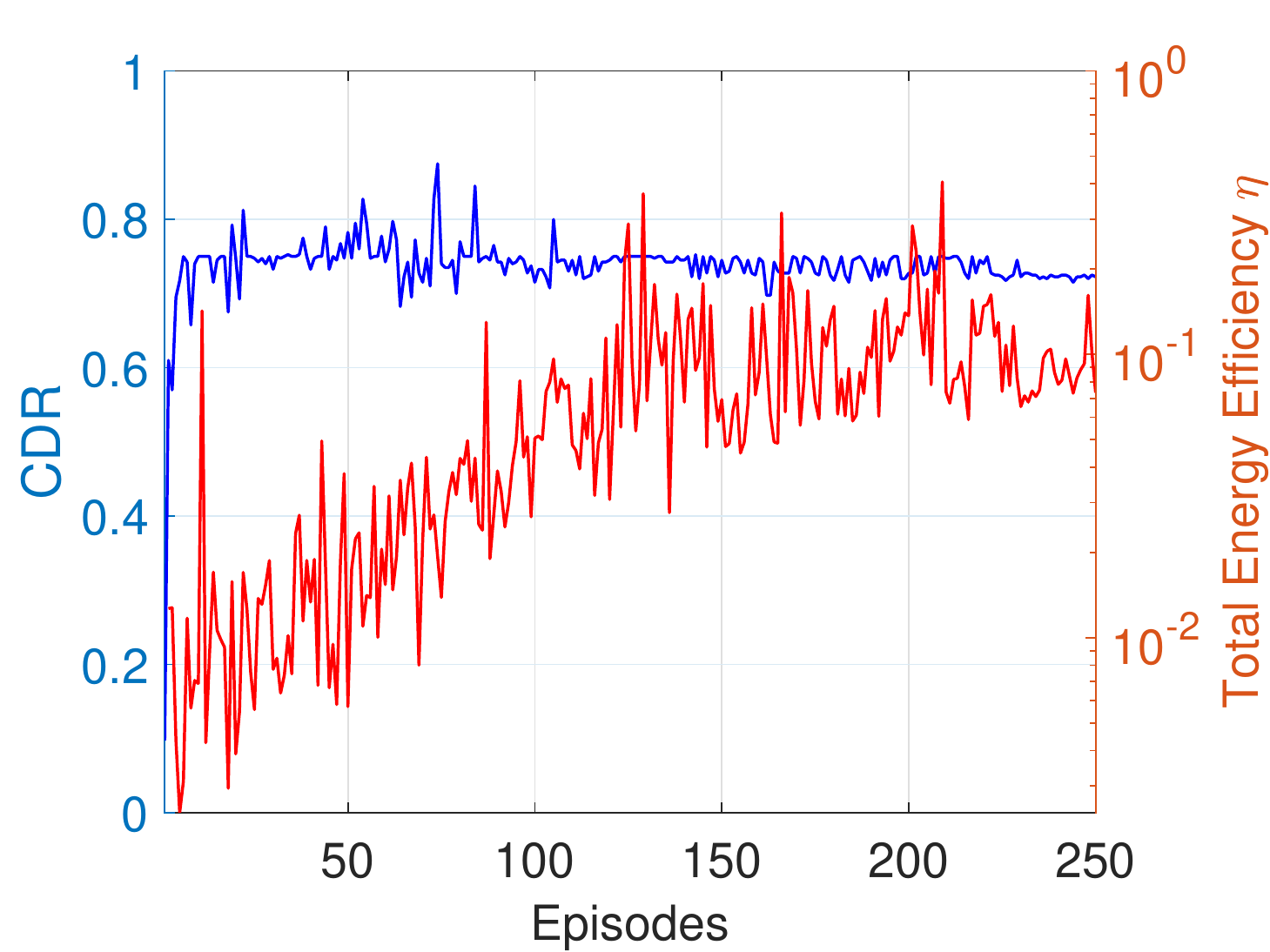}
\label{subfig:edge_cov}}
\end{minipage}
\caption{Deployment of 10 UAVs to provide coverage to static toy-case users in different density scenarios} 
\label{fig: toy cases}
\vspace{-3mm} 
\end{figure*}
 
\begin{itemize}[leftmargin=*]
    \item \textbf{State space}: The state space for Agent $j$ given in line~\ref{Algline:state_start}--\ref{Algline:state_stop} can be expressed as a tuple,~\textlangle{}$x^t : \{0, 1, ..., x_{max}\}, y^t : \{0, 1, ..., y_{max}\},~h^t : \{h_{min},..., h_{max}\},~C_j^t, ~e_j^t,~\frac{C_j^t}{C_j^*},~x^*, ~y^*,~N_d^t, C_z^t, \frac{C^{t}_o}{C^{*}_o},~e_z^t$\textrangle{}, where $\frac{C_j^t}{C_j^*}$ is the ratio of the connectivity score of UAV~$j$ at time-step $t$ to the best connectivity score experienced by the UAV over a series of past encounters. The $x^*$ and $y^*$ are the coordinates where the UAV experienced its best connectivity score. $N_d^t$ is the distance of neighbouring UAVs, $C_z^t$ is the connectivity score of neighbouring UAVs, and $e_z^t$ is the instantaneous energy level of neighbouring UAVs. $\frac{C^{t}_o}{C^{*}_o}$ is the ratio of the connectivity score in UAV~$j$'s neighbourhood at time-step $t$ to the best neighbourhood connectivity score experienced over a series of past encounters. The $C^{t}_o$ is the total number of connected users by UAVs in the neighbourhood. However, the communication cost incurred by the agent per step is bounded by $(U - 1) \times E$ \cite{Ming_MARL_1993}, where $U$ is the number of UAVs within that locality,  $E$ is the number of bits needed to represent each observation by the agent.
    \item \textbf{Action space}: At each time-step~$t \in T$, each UAV executes an action by changing its direction along the coordinates:~$(+x_s,~0)$, $(-x_s,~0)$, $(0,+y_s)$, $(0,-y_s)$, and $(0,~0)$. 
    \item \textbf{Reward}: The goal of the agent is to learn a policy that implicitly maximises the system's EE by jointly maximising the number of connected vehicles while minimising the total UAVs' energy consumption. Hence, we introduce a shared cooperative factor~$\mho$ to shape the reward formulation of each agent $j$ in each time-step $t \in T$ given as,
    \begin{equation}
    \label{eqnreward}
        \mathcal{R}_j^t =
        \begin{cases}
          ~~\mho + \omega + \frac{C_j^t}{C_j^*}, & \text{if}\ C_j^t > ~C_j^{t - 1}\\
          ~~\mho + \omega, & \text{if}\ C_j^{t} = ~C_j^{t - 1}\\
          ~~\mho + \omega - \frac{C_j^t}{C_j^*}, & \text{otherwise,}
        \end{cases}
        \vspace{-1mm}
    \end{equation}
    where ~$C_j^{*}$, $C_j^{t}$,~and~$C_j^{t - 1}$~are the best connectivity score ever experienced by Agent $j$ during the learning cycle, connectivity score in the present and previous time-step, respectively. $\omega~=~\frac{e_j^{t - 1} - e_j^{t}}{e_j^{t} + ~e_j^{t - 1}}$, where $e_j^{t}$ and $e_j^{t - 1}$ are the instantaneous energy consumed by agent $j$ in present and previous time-step, respectively. To enhance cooperation while motivating the agents to pursue a goal of providing coverage to dense areas in the neighbourhood, we compute~$\mho$ as,
    \begin{equation}\label{coop_factor}
    \mho =
    \begin{cases}
      ~+\frac{C^{t}_o}{C^{*}_o}, & \text{if}\ C^{t}_o~> ~C^{t - 1}_o\\
      ~~-\frac{C^{t}_o}{C^{*}_o}, & \text{otherwise.}
    \end{cases}
    \end{equation}
\end{itemize}
\subsection{DDQN Implementation}
\vspace{-1mm}
The neural network (NN) architecture of Agent $j$'s DDQN comprises a 27-dimensional state space~$D_s$ input vector, densely connected to 2 layers with 128 and 64 nodes, with each using a rectified linear unit (ReLU) activation function, leading to an output layer with 5 dimensions~$D_a$ of Q-values corresponding to each possible action. The time complexity of the decentralised double deep Q-network algorithm is approximately $\mathcal{O}\Big(NT \big(D_sW_1 + \sum_{k=1}^{K} W_k W_{k+1}\big)\Big)$~\cite{Tan2022_time_complexity}, where $N$ is the learning episodes, $T$ is the time steps, $K$ is the number of hidden layers of the NN, and $W$ is the number of nodes in each hidden layer. The time complexity of a closely related work and evaluation baseline~\cite{Liu2020UAVdistributed} (MADDPG) is approximately $\mathcal{O}\Big(NT \big(D_sW_1 + \sum_{k=1}^{K} W_k W_{k+1}\big)\Big) + \mathcal{O}\Big(NT\big((D_a + D_s)W_1 + \sum_{k=1}^{K} W_k W_{k+1}\big)\Big)$. Further reading on the implementation and training methodology can be found in \cite[Section IIIC]{omoniwaLetters2022}.
\section{Evaluation, Results and Analysis}
\begin{table}[!t]
\footnotesize
\centering
\caption{Simulation Parameters}
\label{table:parameters}
\begin{tabular}{|l|l|}
  \hline
 \textit{Parameters} & \textit{Value} \\
  \hline \hline
  Software platform/Library & Python/PyTorch\\ 
       Optimiser/Loss function & RMSprop/MSELoss\\ 
       Learning rate/Discount factor & 0.0001/0.95\\       
       Hidden layers/Activation function & 2 (128, 64)/ReLu\\
       Replay memory size/Batch size & 10,000/1024\\  
       Policy/Episodes/maxStep & $\epsilon$-greedy/250/1500\\
       Vehicle speed (SUMO) & [0, 50]~km/h\\ 
       SUMO floating car data output & every 1s $\equiv$ 1 unit time-step\\
       Number of UAVs/Weight & 10/16 kg\\
       UAV speed $V$ & [0, 20]~m/s\\ 
       $\kappa_0$, $\kappa_1$, $\kappa_2$ & 79.85~J/s, 88.63~J/s, 0.018~kg/m\\
       UAVs Altitude/Pathloss exponent~\cite{omoniwaLetters2022}& 120m /2\\
       Nominal battery capacity & 16,000 mAh\\ 
       Maximum transmit power~\cite{omoniwaLetters2022} & 20 dBm\\ 
       Noise power/SINR threshold~\cite{Mozaffari2017UAV} & -130 dBm/5 dB\\
       $B_w$~\cite{omoniwaLetters2022}/ UAV step ($\forall~x_s, y_s$) & 1 MHz / [0--20] m\\
      \hline
 \end{tabular}
\end{table}    

\begin{figure*}[ht]
\begin{minipage}[b]{0.33\linewidth}
\centering
\subfloat[Subfigure 1_ list of figures text][Trajectory of 10 UAVs serving real deployment of vehicles at 10$^{th}$ episode.]{\includegraphics[width=0.9\textwidth]{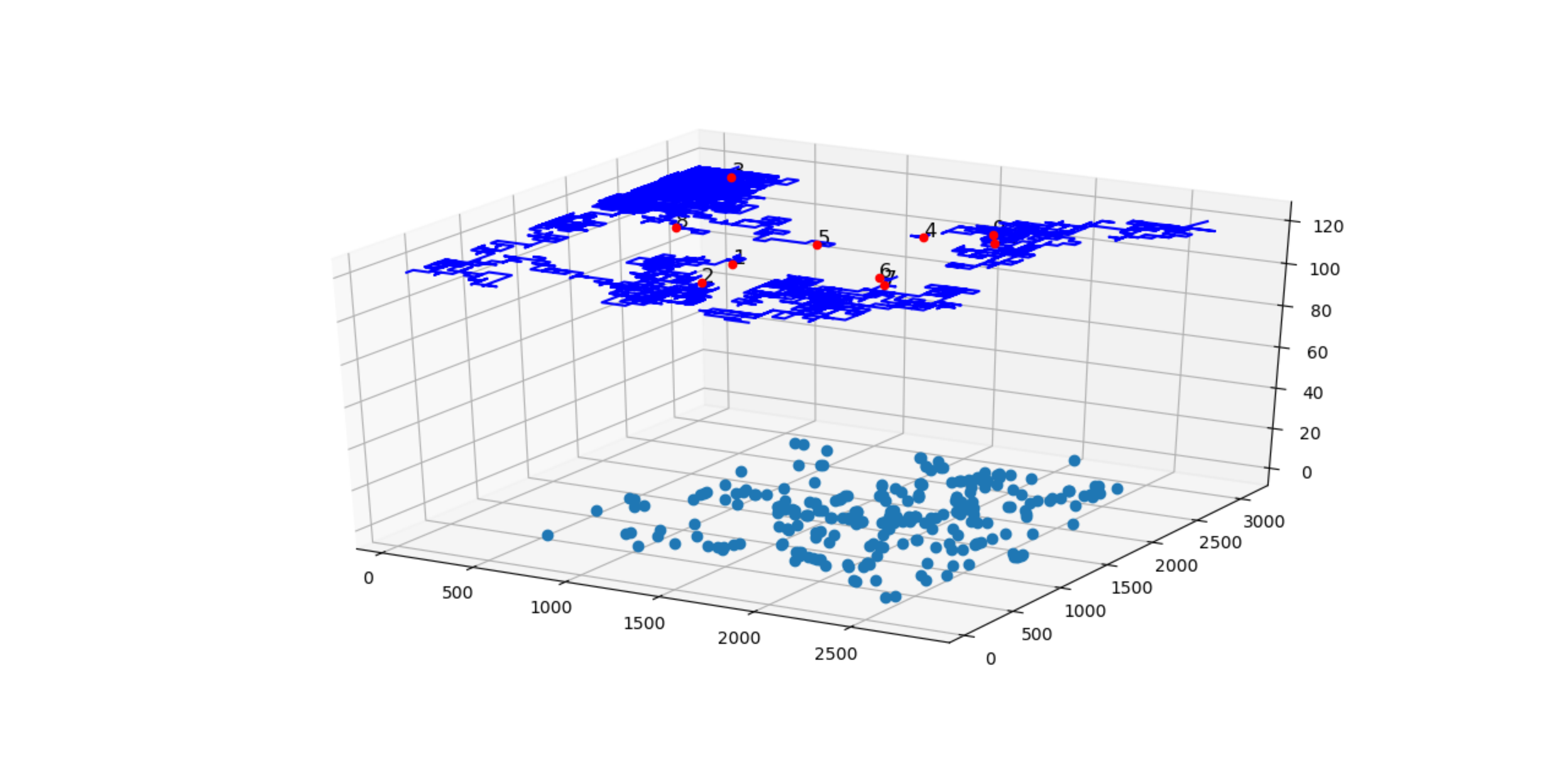}
\label{subfig:sumo_10}}
\end{minipage}
\begin{minipage}[b]{0.33\linewidth}
\centering
\subfloat[Subfigure 1 list of figures text][Trajectory of 10 UAVs serving real deployment of vehicles at 250$^{th}$ episode.]{\includegraphics[width=0.9\textwidth]{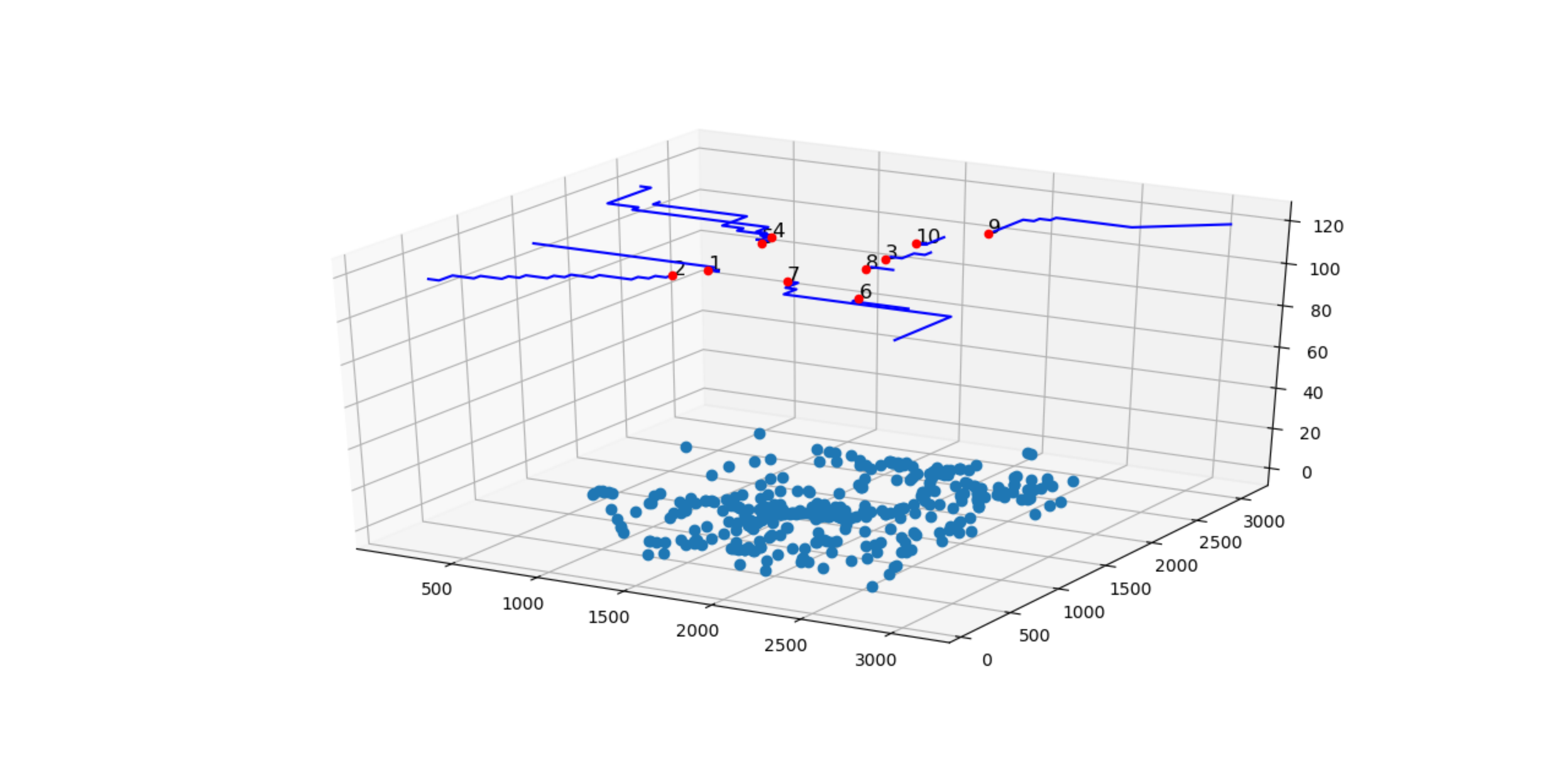}
\label{subfig:sumo_side}}
\end{minipage}
\begin{minipage}[b]{0.33\linewidth}
\centering
\subfloat[Subfigure 2 veh list of figures text][Trajectory of 10 UAVs serving real deployment of vehicles at 250$^{th}$ episode (top view).]{
\includegraphics[width=0.9\textwidth]{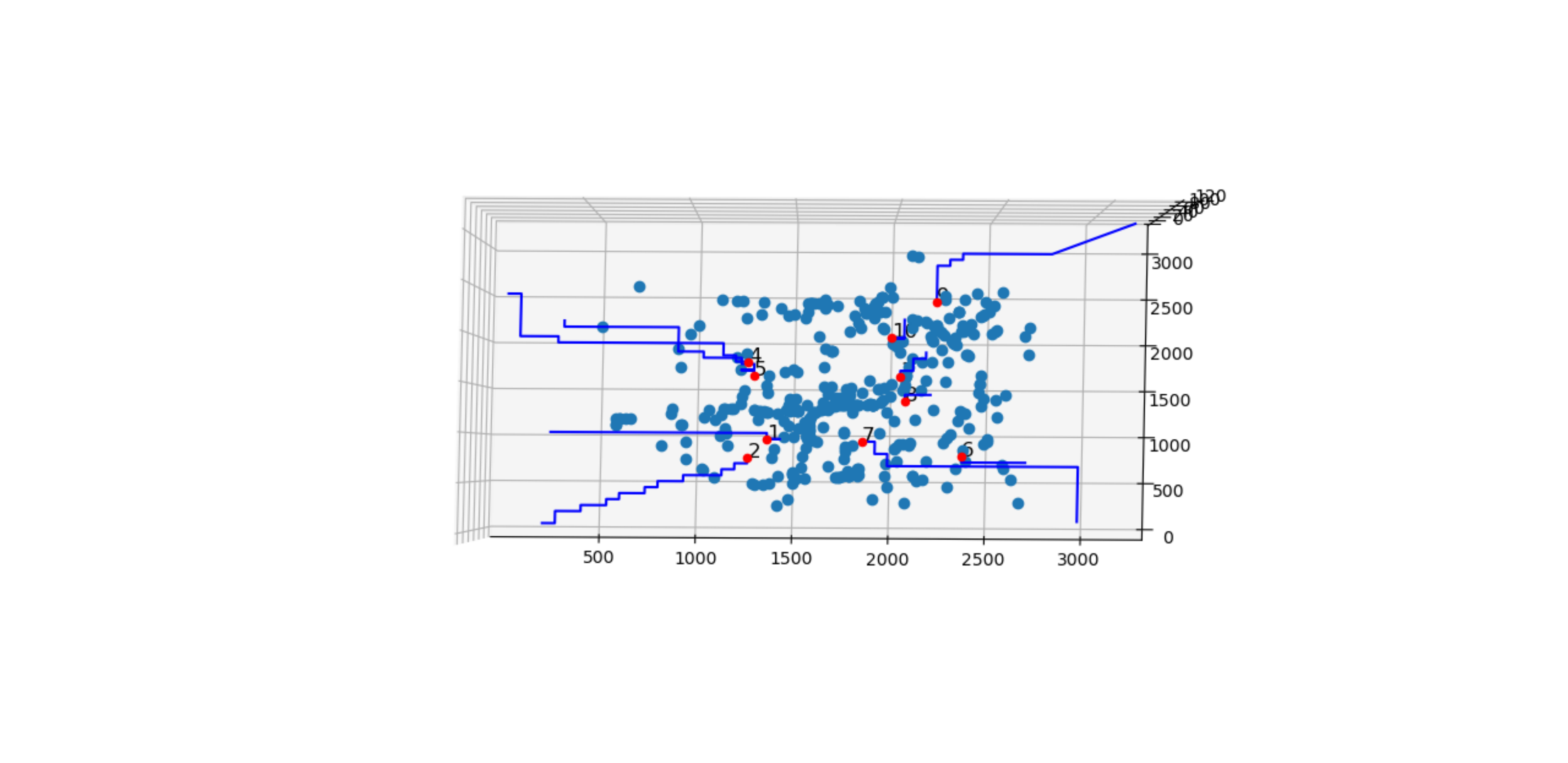}
\label{subfig:sumo_top}}
\end{minipage}
\begin{minipage}[b]{0.32\linewidth}
\centering
\subfloat[Subfigure 2a veh list of figures text][Connected vehicles to deployed vehicles ratio (CDR) vs. episodes.]{
\includegraphics[width=0.9\textwidth]{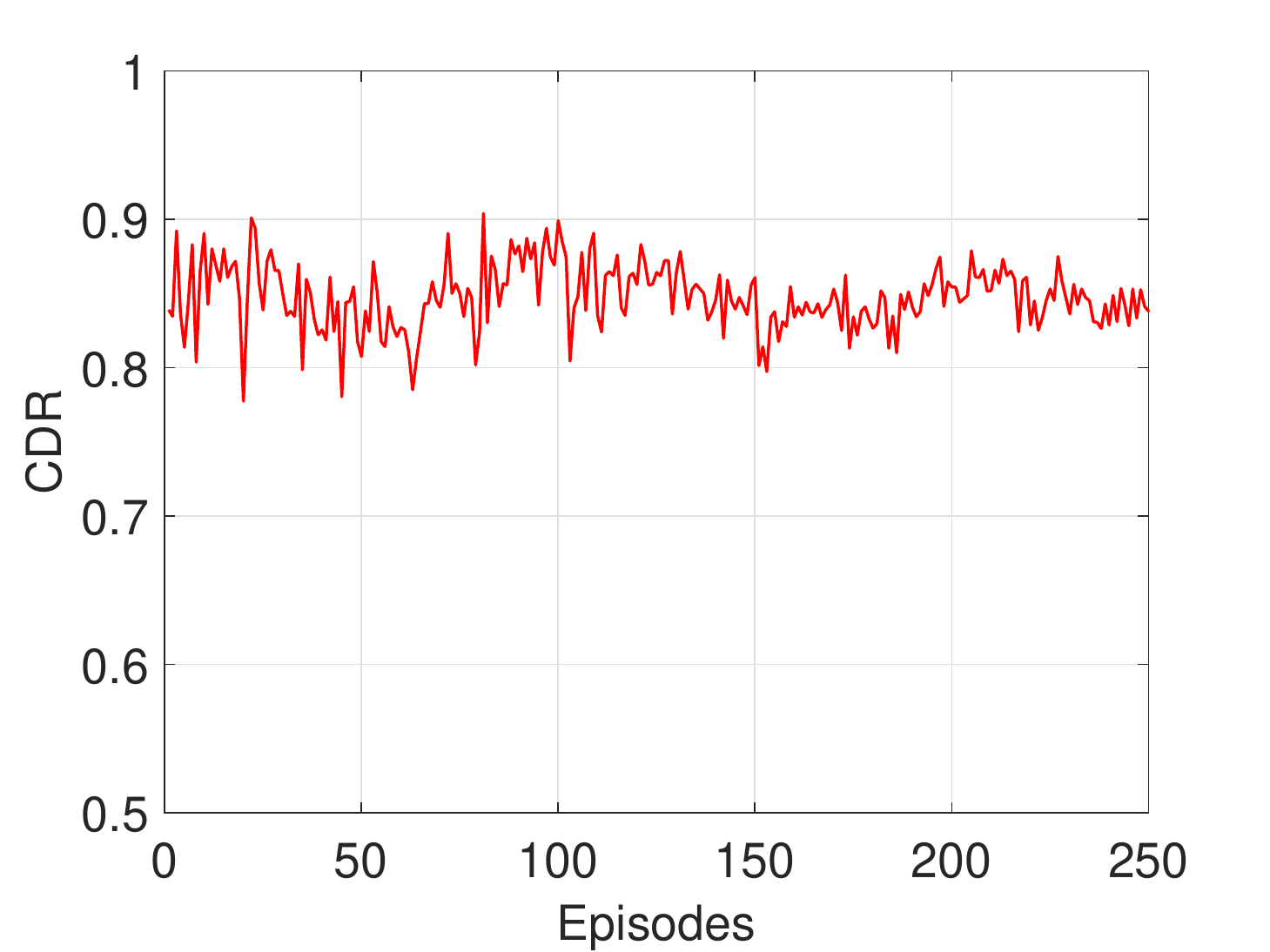}
\label{subfig:CDR_sumo}}
\end{minipage}
\begin{minipage}[b]{0.32\linewidth}
\centering
\subfloat[Subfigure 2c veh list of figures text][Total energy efficiency $\eta$ vs. episodes.]{
\includegraphics[width=0.9\textwidth]{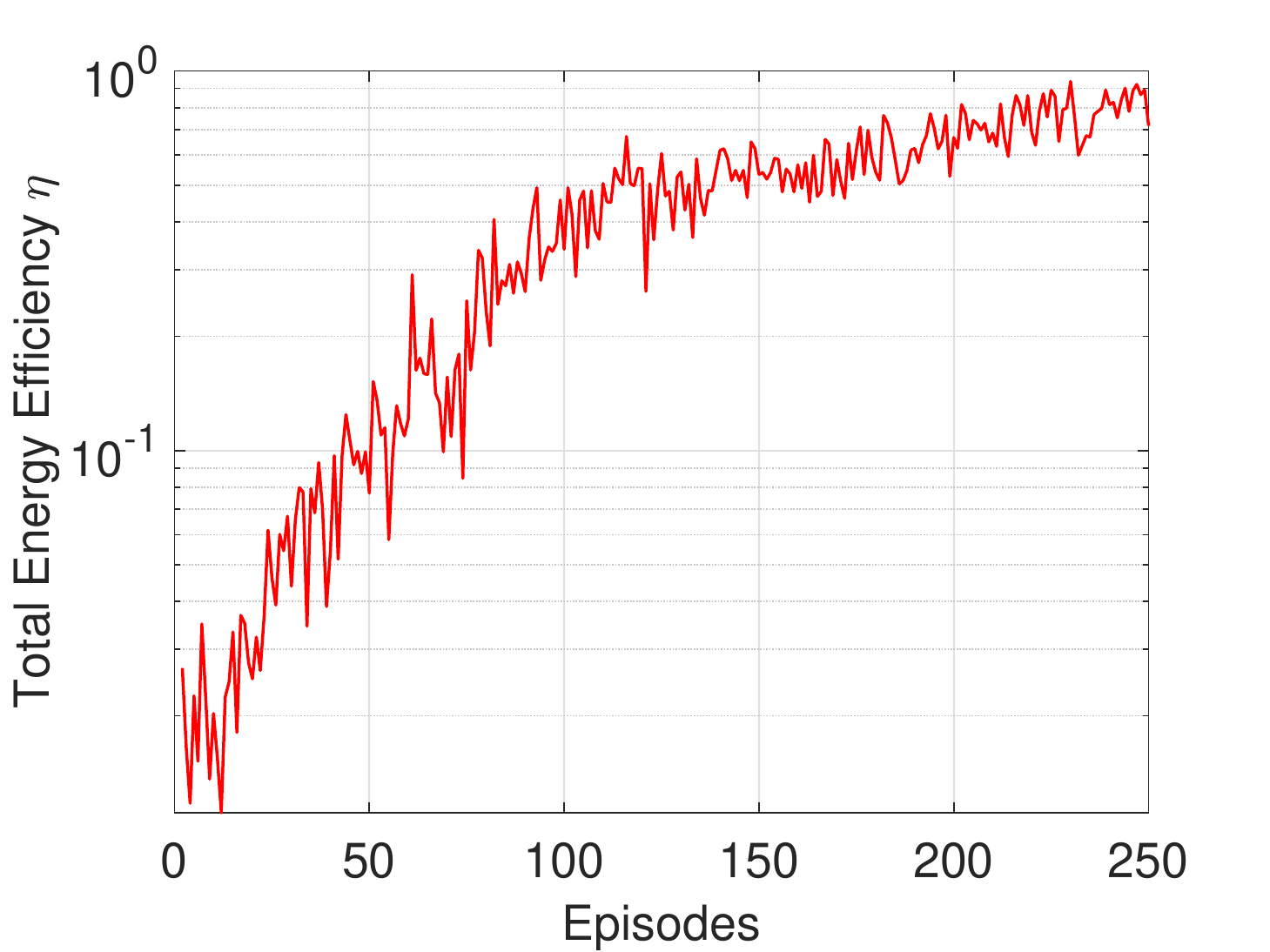}
\label{subfig:ee_sumo}}
\end{minipage}
\begin{minipage}[b]{0.32\linewidth}
\centering
\subfloat[Subfigure 2b veh list of figures text][Total energy consumed vs. episodes.]{
\includegraphics[width=0.9\textwidth]{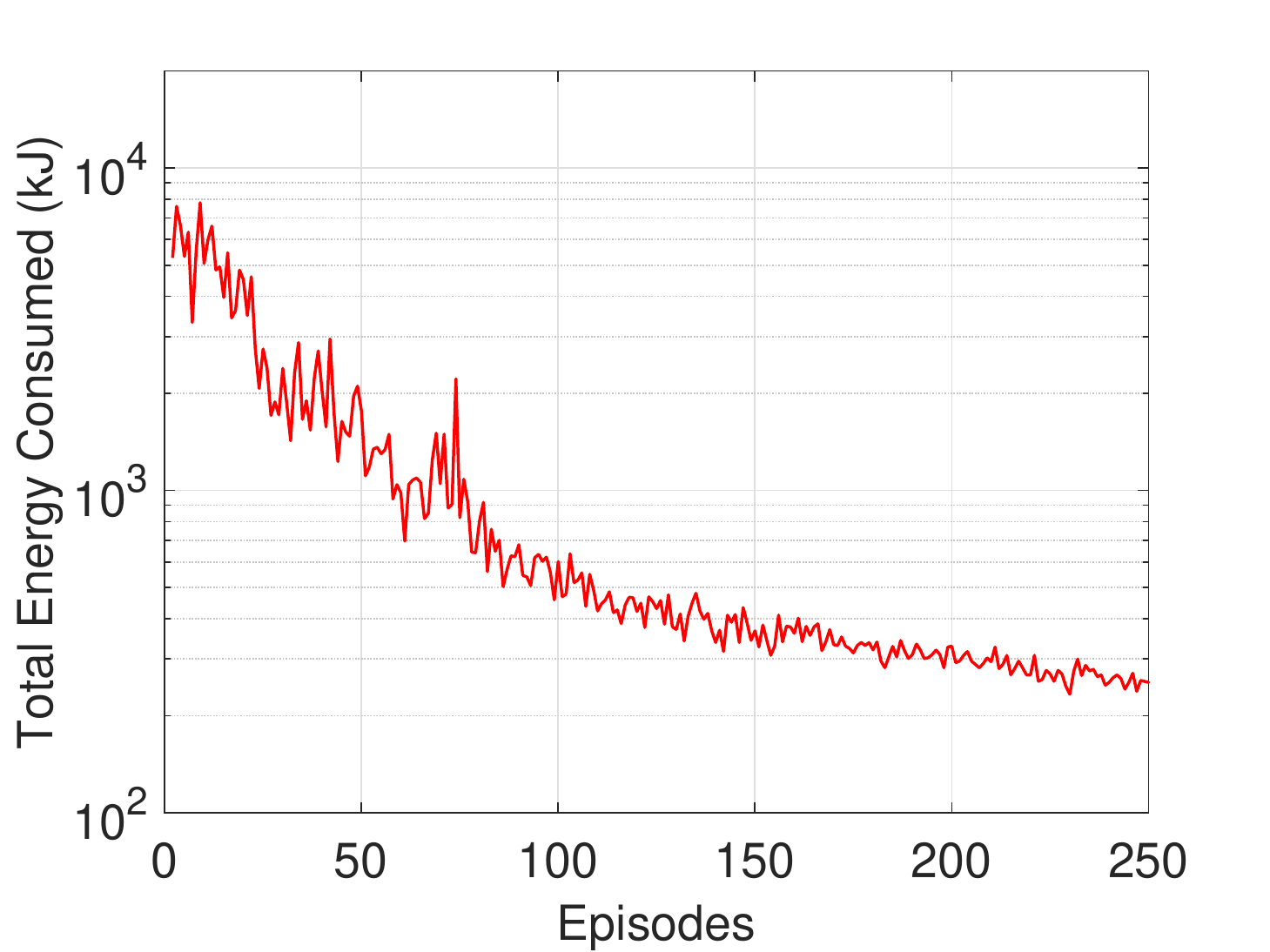}
\label{subfig:energy_sumo}}
\end{minipage}
\begin{minipage}[b]{0.32\linewidth}
\centering
\subfloat[Subfigure 3a veh list of figures text][Connected vehicles to deployed vehicles ratio (CDR) vs. approaches.]{
\includegraphics[width=0.9\textwidth]{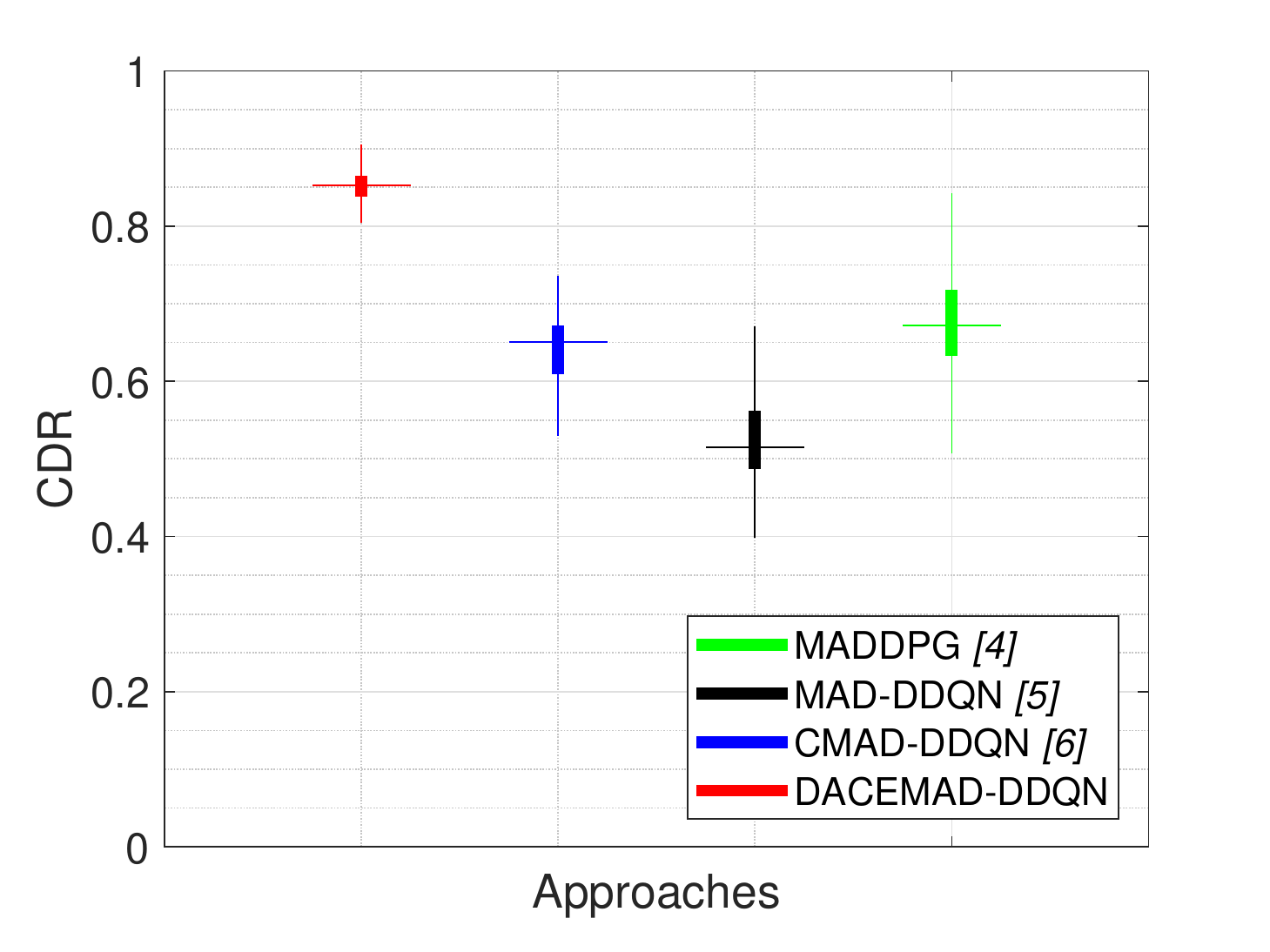}
\label{subfig:CDRVsApproach_sumo}}
\end{minipage}
\hspace{0.2cm}
\begin{minipage}[b]{0.32\linewidth}
\centering
\subfloat[Subfigure 3c veh list of figures text][Total energy efficiency $\eta$ vs. approaches.]{
\includegraphics[width=0.9\textwidth]{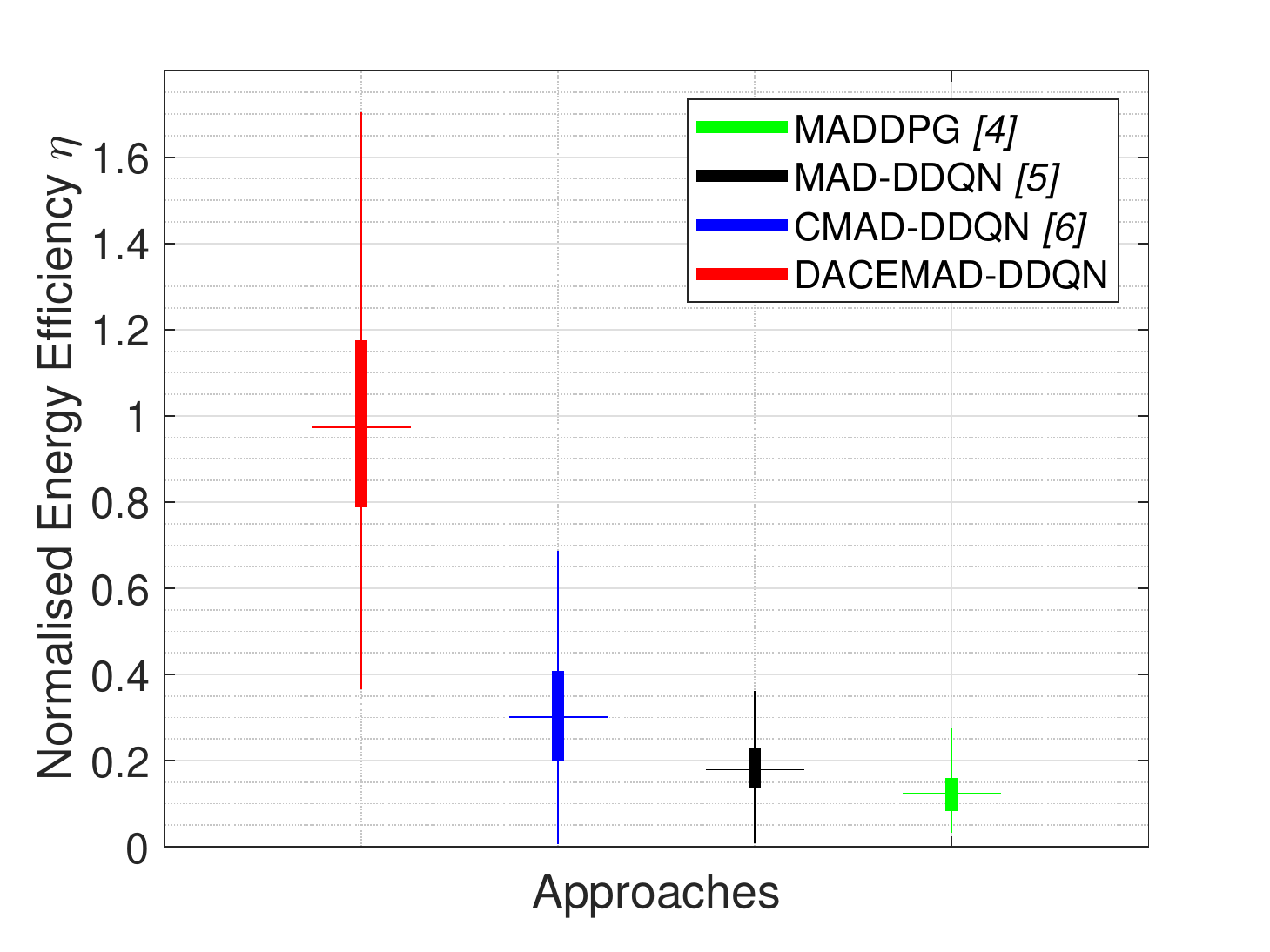}
\label{subfig:eeVsApproach_sumo}}
\end{minipage}
\hspace{0.2cm}
\begin{minipage}[b]{0.32\linewidth}
\centering
\subfloat[Subfigure 3b veh list of figures text][Total energy consumed vs. approaches.]{
\includegraphics[width=0.9\textwidth]{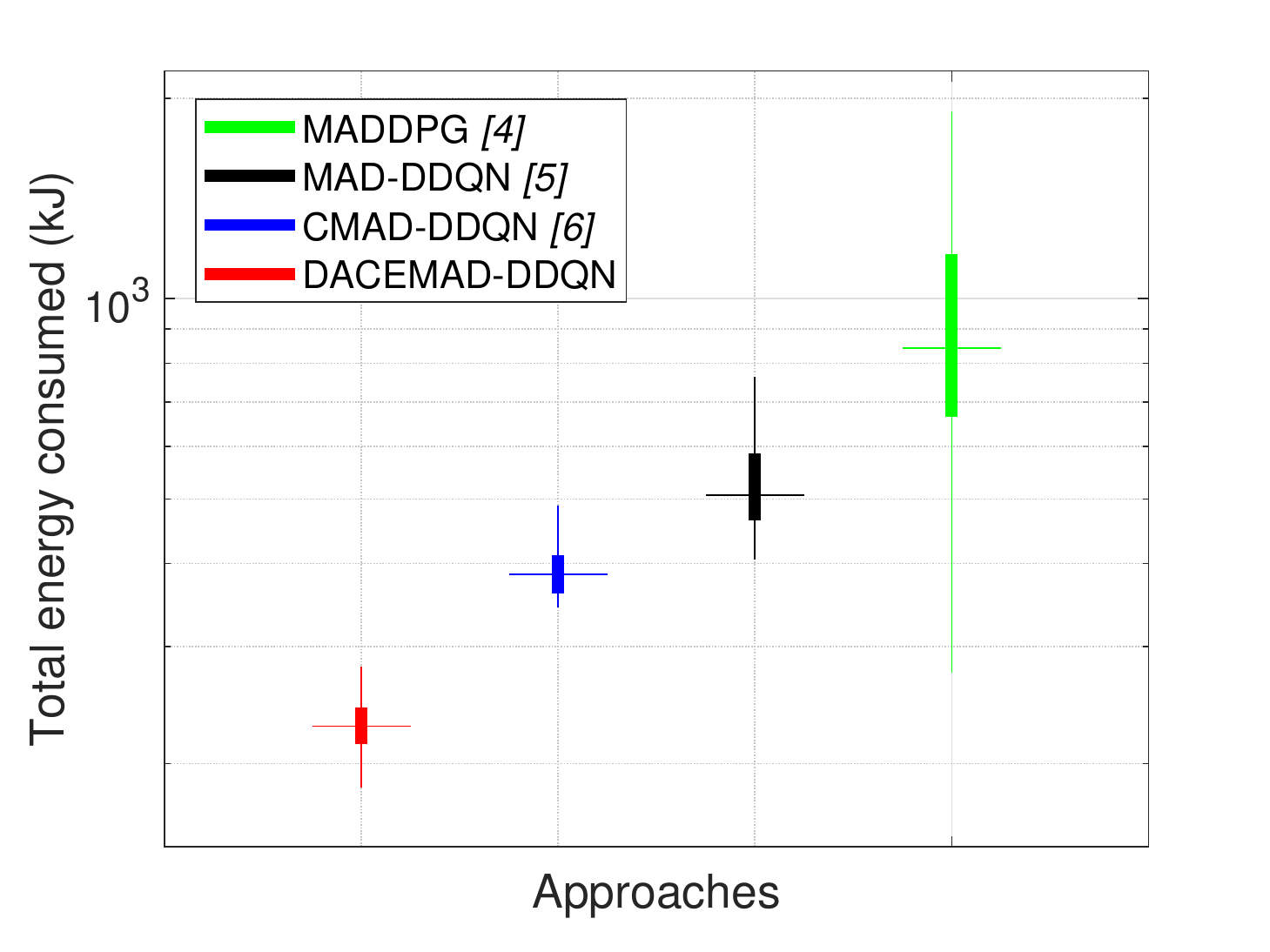}
\label{subfig:energyVsApproach_sumo}}
\end{minipage}
\caption{Agent-controlled UAVs deployed to provide wireless coverage to vehicles using traffic data of the Dublin City Centre generated using SUMO.} 
\label{fig: Real case}
\end{figure*}
Simulation parameters are presented in Table~\ref{table:parameters}. We simulate 10 UAVs to provide wireless coverage to vehicles in a 3000$\times$3000~m$^2$ area. We consider a scenario where the vehicles enter and leave the coverage area. The initial take-off positions of the UAVs are assumed to be known beforehand. To measure the performance of our approach, we consider the connected users to deployed users ratio (CDR), total systems' energy efficiency (EE), and energy consumed as evaluation metrics. 

First, we consider the deployment of 10 UAVs to serve static ground users in 3 different network configurations. The objective of investigating these configurations is to verify the effectiveness of the DACEMAD-DDQN agent-controlled UAVs in serving densely uneven users' distribution. Figures \ref{subfig:circular10}~--~\ref{subfig:edge10} show different distributions of static ground users served by 10 UAVs and their trajectories over a series of time-steps during the 10$^{th}$ learning episode. As expected, we observe a high degree of exploration by the UAVs, leading to random and uncertain policies. However, during the 250$^{th}$ episode, the UAVs' actions were more definite and from Figures \ref{subfig:circular}~--~\ref{subfig:edge_top}, we see that the UAVs are aware of the dense user locations. Figures \ref{subfig:circular_cov}~--~\ref{subfig:edge_cov} show the plots of CDR and total EE against the learning episodes on the different toy scenarios. The results show that the UAVs are capable of coordinating amongst themselves to improve the CDR and the total EE.
 
We leverage real-world traffic data of the Dublin City Centre using SUMO~\cite{Gueriau2020_dublin}. To demonstrate the mobility of the vehicles, we adopt a car-following model called the intelligent driver model to capture traffic phenomena and road user behaviour. We then compare our approach against the following baselines: (\textit{i}) the CMAD--DDQN~\cite{omoniwa_cmad_paper} that considers pedestrians and has no density-aware mechanism, (\textit{ii}) the MAD--DDQN~\cite{omoniwaLetters2022} with no means for direct collaboration and, (\textit{ii}) the MADDPG~\cite{Liu2020UAVdistributed} approach that neglects interference from nearby UAV cells. Figures \ref{subfig:CDR_sumo}, \ref{subfig:ee_sumo} and \ref{subfig:energy_sumo} show the plots of the CDR, total EE and the total energy consumed against the episodes, respectively. As expected, we observed convergence after the 200$^{th}$ episode, which demonstrates the effectiveness of our proposed approach in jointly optimising the total EE and total energy utilisation without degrading the coverage performance.

Results in Figures \ref{subfig:CDRVsApproach_sumo}, \ref{subfig:eeVsApproach_sumo} and \ref{subfig:energyVsApproach_sumo} are obtained from 2000 trained episodes of the agents. Figure \ref{subfig:CDRVsApproach_sumo} shows the plot of CDR versus the approaches considered. The proposed DACEMAD--DDQN approach show robustness and adaptability in providing better connectivity to vehicles compared to the CMAD--DDQN, MAD--DDQN, and MADDPG approaches by approximately 21\%, 33\% and 18\%, respectively. Although the MADDPG approach slightly outperforms the CMAD--DDQN approach by about 3\%, the energy consumed by MADDPG to achieve this performance was significantly higher than that from other approaches.
We normalise the EE values with respect to the mean value of the proposed DACEMAD--DDQN approach. Figure \ref{subfig:eeVsApproach_sumo} shows the plot of the normalised EE versus the approaches considered. From Figure \ref{subfig:eeVsApproach_sumo}, we observe that the DACEMAD--DDQN approach consistently outperforms the CMAD--DDQN, MAD--DDQN, and MADDPG approaches by approximately 65\%, 80\%  and~85\%, respectively. Figure \ref{subfig:energyVsApproach_sumo} shows the plot of the total energy consumed in kiloJoules versus the approaches considered. Our DACEMAD--DDQN approach outperforms baselines in minimising the total energy consumed in the network, while the MADDPG performed worse. Intuitively, direct communication along with the density-aware mechanism of our proposed DACEMAD--DDQN solution enables the UAVs to effectively collaborate to minimise the total energy consumed, while serving highly mobile and densely uneven users' distribution. 
 
\section{Conclusion}
In this work, we propose a Density-Aware Communication-Enabled Multi-Agent Decentralised Double Deep Q-Network (DACEMAD--DDQN) approach suitable in emergencies to optimise the energy efficiency (EE) of a fleet of UAVs serving ground users in a shared, dynamic and interference-limited environment. Here, each deployed UAV collaborates via communication with nearest neighbours to improve the system performance. Furthermore, we consider a density-aware mechanism that enhances the UAVs' ability to serve densely and uneven users' distribution. Specifically, we investigated the deployment of UAVs to serve vehicles using real-traffic data of an urban area. Our DACEMAD--DDQN approach does not rely on a central controller for decision making, and guarantees quick adaptability in both static and vehicular settings. We compared our approach with state-of-the-art decentralised multi-agent reinforcement learning approaches under the same network conditions. The DACEMAD--DDQN approach outperforms the baselines in improving the total systems' EE, while jointly optimising the number of connected vehicles and the total energy consumed by the UAVs under a strict energy budget. Our future work will investigate the performance impact of other cooperative methods that may incur lesser communication overhead. 

\section*{Acknowledgment}
\vspace{-1mm}
This work was supported, in part, by the Science Foundation Ireland (SFI) Grants No. 16/SP/3804 (Enable) and 13/RC/2077\_P2 (CONNECT Phase 2), the National Natural Science Foundation Of China (NSFC) under the SFI-NSFC Partnership Programme Grant Number 17/NSFC/5224.


%

\end{document}